\documentclass[aps,prb,reprint,floatfix,superscriptaddress,amssymb,amsmath]{revtex4-2}
\usepackage{graphicx}
\usepackage{bm}
\usepackage{bbm}
\usepackage{dsfont}
\usepackage{hyperref}
\usepackage{dcolumn}
\usepackage[mathlines]{lineno}
\usepackage{color}
\usepackage{amssymb}

\makeatletter

\newcommand{\Rmnum}[1]{\expandafter\@slowromancap\romannumeral #1@}
\makeatother

\begin{document}

\title{Bosonic Holes in Quadratic Bosonic Systems}

\author{Jia-Ming \surname{Hu}}
\affiliation{School of Physics, Sun Yat-sen University, Guangzhou 510275, China.}
\author{Bo \surname{Wang}}
\affiliation{School of Physics, Sun Yat-sen University, Guangzhou 510275, China.}
\author{Ze-Liang \surname{Xiang}}
\email{xiangzliang@mail.sysu.edu.cn}
\affiliation{School of Physics, Sun Yat-sen University, Guangzhou 510275, China.}
\affiliation{State Key Laboratory of Optoelectronic Materials and Technologies, Sun Yat-sen University, Guangzhou 510275, China}

\begin{abstract}
{Hole degrees of freedom play a central role in the exact solution of quadratic (mean-field) systems. Although a variety of experiments have suggested the existence of bosonic holes, a consistent and complete theory has long been hindered by the ghost problems. Here, we resolve the ghost problem and establish a unified theoretical framework for bosonic holes by introducing the $\mathcal{CPT}$ theory and bosonic particle-hole (PH) transformation. The bosonic analogs of the `Fermi surface' and `Fermi level' are proposed. Furthermore, a PH duality between Hermitian and non-Hermitian quadratic bosonic systems (QBSs) is revealed. In both distinct QBSs, the $\mathcal{C}$-parity is shown to label PH conjugate eigenspaces. Building on this duality, we demonstrate the PH Bogoliubov quasiparticles in $\mathcal{APT}$ symmetric Hamiltonians, investigate the dynamical generation of PH entanglement, and predict Hermitian PH Aharonov-Bohm interference in non-Hermitian QBSs.}
\end{abstract}

\maketitle

\section{INTRODUCTION}

The elementary excitations of quadratic systems are conventionally described by Bogoliubov-de-Gennes (BdG) Hamiltonians~\cite{J.-P.Blaizot1986} within the enlarged PH spaces. Such excitations play a significant role in the very few many-body problems that admit exact solutions, where the fermionic and bosonic Bogoliubov modes underpin the mean-field theories of superconductivity~\cite{J.Bardeen1957,N.Boboliubov1958} and superfluidity~\cite{N.Bogoliubov1971}, respectively. 

While fermionic Bogoliubov modes are widely understood as superpositions of electrons and electron holes, the nature of bosonic Bogoliubov modes remains debated, owing to the unclear physical meaning of bosonic holes. 
One viewpoint identifies bosonic Bogoliubov modes, which involve bosonic creation and annihilation operators, as PH superpositions~\cite{A.McDonald2018}. In the alternative viewpoint, although the creation operator of a particle can still be regarded as the annihilation operator of a hole, the PH coupling will drive the bosonic systems into a complex spectrum regime~\cite{B.Wu2001}, where Bogoliubov modes and superfluidity break down. This complex spectrum originates from the negative norms of bosonic holes and the broken $\mathcal{PT}$ symmetry in the bosonic BdG description~\cite{B.Wu2001,B.Wu2003,B.Galilo2015,E.Verhagen2022}. Negative-norm states have long been debated in various areas of physics~\cite{W.Pauli1949,K.S.Stelle1977,M.SHER1989,S.W.Hawking2002}: historically called ``ghosts,'' they are regarded as unphysical because they violate basic postulates of quantum mechanics. However, the hole branches and the complex spectra in QBSs have been observed experimentally in recent years~\cite{E.Verhagen2022,N.R.Bernier2018}. Consequently, an accurate description of bosonic holes and the resolution of the associated ghost problem constitute an indispensable part of a comprehensive study of QBSs~\cite{B.Wu2003}.

Meanwhile, in non-Hermitian systems, the $\mathcal{PT}$ symmetry has been shown to yield anomalous real energy spectra~\cite{C.M.Bender1998}, which has attracted widespread attention over the past two decades~\cite{A.Mandilara2002,A.Mostafazadeh2003,C.M.Bender2007,V.V.Konotop2016,R.El-Ganainy2018,Ş.K.Özdemir2019}. Remarkably, such systems contain exactly half of their eigenstates with negative $\mathcal{PT}$-norm, a feature that, as noted by Bender et al., suggests an underlying charge-conjugation symmetry $\mathcal{C}$~\cite{C.M.Bender2002}.
The eigenvalues of $\mathcal{C}$ coincide with the $\mathcal{PT}$-norms of the corresponding states, called $\mathcal{C}$-parity. 
As a complement to $\mathcal{PT}$ symmetry, the ghost states can be interpreted as normal physical states with positive $\mathcal{CPT}$-norms. Importantly, $\mathcal{C}$-parity distinguishes ``particle'' from ``anti-particle'' states~\cite{C.M.Bender2015}, which aligns conceptually with the notion of ``hole'' in condensed-matter systems. This striking correspondence points to a deep connection between the $\mathcal{C}$ symmetry of $\mathcal{PT}$ symmetric non-Hermitian systems and the hole degrees of freedom in Hermitian QBSs within the $\mathcal{PT}$ symmetric BdG framework.

In this paper, we develop a second-quantization theory of bosonic holes and uncover the PH duality between non-Hermitian and Hermitian QBSs that share the same physical properties. By introducing the $\mathcal{CPT}$ theory in Hermitian QBSs, we have resolved the ghost problem associated with hole-like states and derived the form of quasi-holes, whose spectra have already been observed experimentally~\cite{E.Verhagen2022}. We further show that the hole (quasihole) Fock states are occupied by negative particle (quasiparticle) numbers, which is the defining characteristic that identifies hole degrees of freedom and gives rise to ghost problems within the second-quantization description. To address this, we propose a non-unitary PH transformation to construct exact positive-norm Fock states for bosonic holes and define a bosonic analog of `Fermi surface' to ensure positive particle numbers in displaced hole states. The negative energies, arising from opposite PH energy spectra, can be solved via a bosonic `Fermi level' analog. 

Then, we take the well-known $\mathcal{APT}$-symmetric Hamiltonian~\cite{F.Yang2017,Y.Yang2020} as an example, showing that the `particle' and `anti-particle' single-particle states in non-Hermitian QBSs give rise to elementary excitations in PH conjugate spaces. These results are consistent with those in Hermitian weakly interacting Bose gases~\cite{N.Bogoliubov1971,S.Sachdev2023} under a local PH transformation, reflecting the PH duality between Hermitian and non-Hermitian QBSs. Beyond $\mathcal{PT}$ physics, PH duality predicts Hermitian BdG Hamiltonians with fully real spectra in non-Hermitian QBSs, exemplified by a recent work~\cite{A.A.Clerk2023}. Moreover, we formulate a gauge theory in the PH space to derive the Hermitian PH Aharonov-Bohm (AB) interference and the resulting chiral flows in non-Hermitian QBSs.

Our paper is organized as follows. Section~\Rmnum{2} introduces the explicit form of bosonic holes and discusses their fundamental properties. In Sec.~\Rmnum{3}, we employ the $\mathcal{CPT}$ inner product to resolve the ghost problem in the bosonic BdG Hamiltonian, leading to the identification of quasi-holes and PH superpositions in QBSs. Section~\Rmnum{4} formulates the second quantization theory of bosonic holes, including the Fock states and the physical picture. In Sec.~\Rmnum{5}, we reveal the PH duality between Hermitian and non-Hermitian QBSs that exhibit the same physical behavior in different representations. Section~\Rmnum{6} presents an experimentally feasible non‑Hermitian system that explicitly demonstrates hole degrees of freedom. The advantages of our framework for studying system dynamics, compared with conventional approaches, are illustrated in Sec.~\Rmnum{7}. In Sec.~\Rmnum{8}, we predict an emergent Hermitian PH Aharonov‑Bohm effect in a non‑Hermitian system. The main results of this work are summarized in Sec.~\Rmnum{9}.


\section{Hole degrees of freedom in BdG description}

We start with a general Hermitian quadratic bosonic Hamiltonian 
\begin{equation}
\hat{H}=\sum_{i,j}\hat{a}_i^\dagger\mathcal{A}_{ij}\hat{a}_j+\frac{1}{2}(\hat{a}_i^\dagger\mathcal{B}_{ij}\hat{a}_j^\dagger+\hat{a}_i\mathcal{B}_{ij}^\ast \hat{a}_j). 
\end{equation}
The Hermiticity of $\hat{H}$ requires $\mathcal{A}_{ij}=\mathcal{A}_{ji}^\ast$, while the bosonic commutation relation allows the pairing matrix to be chosen symmetric, $\mathcal{B}_{ij}=\mathcal{B}_{ji}$. Introducing the Nambu array 
\begin{equation}
\hat{\alpha}={(\hat{a}_1\cdots \hat{a}_N, \hat{a}_1^\dagger\cdots \hat{a}_N^\dagger)}^T,
\end{equation}
the Hamiltonian takes the compact form 
\begin{equation}
\hat{H}=\frac{1}{2}\hat{\alpha}^\dagger H\hat{\alpha},
\end{equation}
and the Heisenberg equation reads 
\begin{equation}
i\partial_t\hat{\alpha}=\mathcal{H}\hat{\alpha}.
\end{equation}
We write the quadratic form in terms of the matrices
\begin{equation}
\begin{split}
&H=\left(\begin{array}{cc}
\mathcal{A} & \mathcal{B}\\
\mathcal{B}^\ast& \mathcal{A}^\ast 
\end{array}\right),\\
&\mathcal{H}=\tau_3H=\left(\begin{array}{cc}
\mathcal{A} & \mathcal{B}\\
-\mathcal{B}^\ast& -\mathcal{A}^\ast 
\end{array}\right),
\end{split}
\end{equation}
where $\tau_i=\sigma_i\otimes I_N$ from the Pauli matrices $\sigma_i$ ($i=1, 2, 3$). The difference between $H$ and $\mathcal{H}$ becomes apparent from 
\begin{equation}
    \hat{H}=1/2\hat{\alpha}^\dagger H\hat{\alpha}=1/2\hat{\bar{\alpha}}^\dagger \mathcal{H}\hat{\alpha},
\end{equation}
with 
\begin{equation}
    \hat{\alpha}^\dagger=(\hat{a}_1^
\dagger,\cdots,\hat{a}_N^\dagger,\hat{a}_1,\cdots,\hat{a}_N)
\end{equation}
and
\begin{equation}
    \hat{\bar{\alpha}}^\dagger=\hat{\alpha}^\dagger\tau_3=(\hat{a}_1^
\dagger,\cdots,\hat{a}_N^\dagger,-\hat{a}_1,\cdots,-\hat{a}_N).
\end{equation}
In the existing literature, the operators $\hat{a}_i^\dagger$ in $\hat{\alpha}$ are often formally interpreted as annihilation operators of bosonic holes. However, it is important to note that the commutation relations
\begin{equation}
[\hat{a}_i^\dagger,\hat{a}_{j}]=-\delta_{ij},\quad [\hat{a}_i^\dagger,-\hat{a}_{j}]\equiv[\hat{h}_i, \hat{\bar{h}}_j^\dagger]=\delta_{ij}, 
\end{equation}
reflect that the canonical bosonic commutation relations are naturally encoded in the basis of $\mathcal{H}$, instead of $H$, where the operators $-\hat{a}_i$ in $\hat{\bar{\alpha}}^\dagger$, rather than $\hat{a}_i$ in $\hat{\alpha}^\dagger$, should be properly identified as the creation operators of bosonic holes. Therefore, the spectrum of the systems can be obtained by directly diagonalizing $\mathcal{H}$, i.e., the bosonic BdG Hamiltonian, while the matrix $H$ requires symplectic diagonalization~\cite{S.Lieu2018,L.Viola2020}.

The operators of bosonic holes exhibit two remarkable properties. First, bosonic holes are pseudo-bosonic modes, meaning that the transpose-conjugate of the annihilation operator is not equivalent to the creation operator~\cite{R.Rossignoli2005,D.A.Trifonov2009,R.Rossignoli2017}, i.e., 
\begin{equation}
\hat{\bar{h}}_i^\dagger\ne \hat{h}_i^\dagger.
\end{equation}
Second, they correspond to negative-particle excitations, implied in the bosonic commutation relation 
\begin{equation}
[\hat{a}^\dagger,\hat{a}]=\hat{a}^\dagger\hat{a}+\hat{\bar{h}}^\dagger\hat{h}=-1.
\end{equation}

The first property underlies the non-Hermitian nature of the bosonic BdG Hamiltonian $\mathcal{H}$ (satisfying $\mathcal{H}^\dagger\ne \mathcal{H}$) for Hermitian QBSs described by $\hat{H}$. It then follows that any superpositions of bosonic particles and holes must be pseudo-bosonic modes. Together with the negative-particle property, these provide clear criteria for identifying whether a quasiparticle is a PH superposition. In this sense, the usual Bogoliubov modes, which are canonical bosonic modes with positive particle expectation in each component, cannot be regarded as superpositions of bosonic particles and holes.


\section{Bosonic holes in elementary excitations of QBSs}

To clarify how bosonic holes contribute to the elementary excitations of Hermitian QBSs, we begin by identifying the intrinsic symmetries of the bosonic BdG Hamiltonian. Similar to the fermionic case, $\mathcal{H}$ inherently respects PH symmetry 
\begin{equation}
C\mathcal{H}C=-\mathcal{H},
\end{equation}
where $C=\tau_1\mathcal{K}$ with $\mathcal{K}$ being complex conjugation. Differently, instead of Hermiticity, the PH-symmetric structure brings the generalized $\mathcal{PT}$ symmetry 
\begin{equation}
\mathcal{PT}\mathcal{H}\mathcal{PT}=\mathcal{H}^T,
\end{equation}
where $\mathcal{P}=\tau_3$ and $\mathcal{T}=\mathcal{K}$. The interest in generalized $\mathcal{PT}$ symmetry has arisen from its role in providing real energy spectra in non-Hermitian systems over the past decades. Its emergence within Hermitian QBSs has recently attracted growing attention, given its potential to enable the study of dissipation-free non-Hermitian dynamics~\cite{E.Verhagen2022,A.McDonald2018,A.A.Clerk2019,Z.X.Zhou2020,X.-Y.Lv2023,C.C.Wanjura2023,E.Verhagen2024}.

If $\vert \psi_n\rangle$ is an eigenvector of $\mathcal{H}$ with eigenvalue $\epsilon_n$, the PH symmetry ensures that, $C\vert \psi_n\rangle$ is an eigenvector of $\mathcal{H}$ with eigenvalue $-\epsilon_n^\ast$, i.e., 
\begin{equation}
\mathcal{H}C\vert \psi_n\rangle=C(-\mathcal{H})\vert \psi_n\rangle=C(-\epsilon_n)\vert \psi_n\rangle=-\epsilon_n^\ast C\vert \psi_n\rangle. 
\end{equation}
In addition, the $\mathcal{PT}$ symmetry guarantees that $\epsilon_n^\ast$ is also an eigenvalue of $\mathcal{H}$. Consequently, the eigenvalues of $\mathcal{H}$ come in quartets $\{\epsilon_n, \epsilon_n^\ast, -\epsilon_n, -\epsilon_n^\ast\}$, with degeneracies occurring when $\epsilon_n$ is real or imaginary. 

\subsection{$\mathcal{PT}$-symmetry unbroken regime}

When the system is $\mathcal{PT}$-symmetry unbroken, it possesses real spectra ($\epsilon_n\in \mathbb{R}$). In this regime, the eigenvalues can be divided into two groups $\{\epsilon_n, -\epsilon_n\}$, and the corresponding eigenvectors $\{\vert\psi_n\rangle,C\vert\psi_n\rangle=\vert\tilde{\psi}_n\rangle\}$ are referred to as particle-like and hole-like states, respectively. Owing to the $\mathcal{PT}$ symmetry of $\mathcal{H}$, the system admits a $\mathcal{PT}$-inner product distinct from that in conventional quantum mechanics, given by 
\begin{equation}{(\mathcal{PT}\vert\psi_n\rangle)}^T\vert\psi_m\rangle=\langle \psi_n\vert\tau_3\vert\psi_m\rangle=\delta_{nm},
\end{equation}
which follows from 
\begin{equation}
\mathcal{H}^\dagger\mathcal{P}\vert \psi_n\rangle=\mathcal{P}\mathcal{P}\mathcal{H}^\dagger\mathcal{P}\vert \psi_n\rangle=\mathcal{P}\mathcal{H}\vert \psi_n\rangle=\epsilon_n\mathcal{P}\vert \psi_n\rangle \label{Eq7}.
\end{equation}

In conventional non-Hermitian $\mathcal{PT}$-symmetric systems, it is well known that half of the eigenstates carry negative $\mathcal{PT}$-norms, referred to as antiparticle states, and resolved by introducing the hidden charge-conjugation symmetry $\mathcal{C}$ that commutes with $\mathcal{PT}$~\cite{C.M.Bender2002,C.M.Bender2015}. In the present Hermitian $\mathcal{PT}$-symmetric system, the hole-like eigenstates, which are exactly half of all eigenvectors, play the same role as antiparticle states, exhibiting negative $\mathcal{PT}$-norms:
\begin{equation}
{(\mathcal{PT}\vert\tilde{\psi}_n\rangle)}^T\vert\tilde{\psi}_m\rangle=\langle \psi_n\vert C\tau_3C\vert\psi_m\rangle=-\delta_{nm}.
\end{equation}

The opposite $\mathcal{PT}$-norm signs of particle-like and hole-like states stem from the intrinsic redundancy of quasiparticles and quasiholes, meaning that either set alone is sufficient to describe the full dynamics. Historically, bosonic holes were regarded as non-physical degrees of freedom and not independent from particles~\cite{B.Wu2001}. However, the recent experiment has observed both quasiparticle and quasihole branches in the spectrum~\cite{E.Verhagen2022}, definitively confirming the physical existence of quasihole states. 

Here, we follow the standard $\mathcal{CPT}$-inner-product approach used in non-Hermitian systems to address the ghost problem associated with hole-like states and further investigate the structure of quasihole excitations. Distinct from the PH symmetry operator $C$, which maps particle-like states to hole-like states with opposite $\mathcal{PT}$-norms, we introduce a charge-conjugation operator $\mathcal{C}$ that labels the signs of $\mathcal{PT}$-norms, i.e., 
\begin{equation}
\mathcal{C}\vert\psi_n\rangle=\vert\psi_n\rangle,\quad \mathcal{C}\vert\tilde{\psi}_n\rangle=-\vert\tilde{\psi}_n\rangle.
\end{equation}
This means that, in addition to $\mathcal{PT}$ symmetry, the particle-like and hole-like states possess an additional hidden $\mathcal{C}$ symmetry. The eigenvalues of $\mathcal{C}$ therefore represent the $\mathcal{C}$ parity of each eigenstate. Such a charge-conjugation operator can take the form 
\begin{equation}
    \mathcal{C}=\vert\psi_n\rangle\langle\psi_n\vert\mathcal{P}-\vert\tilde{\psi}_n\rangle\langle\tilde{\psi}_n\vert(-\mathcal{P}).
\end{equation}
Under the $\mathcal{CPT}$-inner product, both particle-like and hole-like states acquire positive norms:
\begin{equation}
\begin{split}
{(\mathcal{PT}\vert\psi_n\rangle)}^T\mathcal{C}\vert\psi_m\rangle=&\langle \psi_n\vert\tau_3\vert\psi_m\rangle=\delta_{nm},\\
{(\mathcal{PT}\vert\tilde{\psi}_n\rangle)}^T\mathcal{C}\vert\tilde{\psi}_m\rangle=&\langle \tilde{\psi}_n\vert-\tau_3\vert\tilde{\psi}_m\rangle=\delta_{nm},
\end{split}
\end{equation}
from which it follows that 
\begin{equation}
\langle\psi_n\vert\tau_3\mathcal{H}\vert\psi_n\rangle=\epsilon_n, \quad \langle\tilde{\psi}_n\vert-\tau_3\mathcal{H}\vert\tilde{\psi}_n\rangle=-\epsilon_n.
\end{equation}
With this $\mathcal{CPT}$-inner product, the identity operator can be defined as 
\begin{equation}\mathbb{I}_{2N}=\sum_{n=1}^{N}\vert\psi_n\rangle\langle\psi_n\vert\tau_3-\vert\tilde{\psi}_n\rangle\langle\tilde{\psi}_n\vert\tau_3.
\end{equation}
Inserting the identity, $\mathcal{H}$ can be diagonalized as
\begin{equation}
\mathcal{H}=\sum_{n=1}^N\epsilon_n\vert\psi_n\rangle\langle\psi_n\vert\tau_3-\epsilon_n\vert\tilde{\psi}_n\rangle\langle\tilde{\psi}_n\vert(-\tau_3).
\end{equation}

Then the quasiparticles and quasiholes can be obtained from particle-like states and hole-like states, respectively, by substituting $\mathcal{H}$ into $\hat{H}=\frac{1}{2}\hat{\alpha}^\dagger\tau_3\mathcal{H}\hat{\alpha}$, yielding 
\begin{equation}
\begin{split}
\hat{H}=&\frac{1}{2}\sum_{n=1}^N(\epsilon_n\hat{\alpha}^\dagger\tau_3\vert\psi_n\rangle\langle\psi_n\vert\tau_3\hat{\alpha}-\epsilon_n\hat{\alpha}^\dagger\tau_3\vert\tilde{\psi}_n\rangle\langle\tilde{\psi}_n\vert(-\tau_3)\hat{\alpha})\\
=&\frac{1}{2}\sum_{n=1}^N(\epsilon_n\hat{\psi}_n^\dagger\hat{\psi}_n-\epsilon_n\hat{\bar{\phi}}_n^\dagger\hat{\phi}_n),
\end{split}
\end{equation}
where $\hat{\psi}_n^\dagger=\hat{\alpha}^\dagger\tau_3\vert\psi_n\rangle$ and $\hat{\psi}_n=\langle\psi_n\vert\tau_3\hat{\alpha}$ are the creation and annihilation operators of quasiparticles, while $\hat{\bar{\phi}}_n^\dagger=\hat{\alpha}^\dagger\tau_3\vert\tilde{\psi}_n\rangle$ and $\hat{\phi}_n=\langle\tilde{\psi}_n\vert-\tau_3\hat{\alpha}$ are the creation and annihilation operators of quasiholes. The canonical bosonic nature of quasiparticles reflects that they are only composed of particles, while quasiholes are pseudo-bosonic modes that consist solely of bosonic holes due to the PH symmetry. Note that the redundancy of quasiparticles and quasiholes can be seen clearly in the following equations
\begin{equation}
    \begin{split}
\hat{\bar{\phi}}_n^\dagger=&\hat{\alpha}^\dagger\tau_3\vert\tilde{\psi}_n\rangle=-\hat{\alpha}^\dagger\tau_1\tau_3\mathcal{K}\vert\psi_n\rangle=-\hat{\psi}_n,\\
\hat{\phi}_n=&\langle\tilde{\psi}_n\vert-\tau_3\hat{\alpha}=\langle\psi_n\vert\mathcal{K}\tau_3\tau_1\hat{\alpha}=\hat{\psi}_n^\dagger.
    \end{split}
\end{equation}
The negative sign in the $\mathcal{CPT}$-inner product (or the negative $\mathcal{PT}$-norms) of the hole-like states precisely makes the creation operators of the quasiholes not equal to the transposed-conjugate of the annihilation operators, but have a negative sign difference, which ensures the bosonic commutation relations
\begin{equation}
    [\hat{\psi}_n,\hat{\psi}_m^\dagger]=[\hat{\phi}_n,\hat{\bar{\phi}}_m^\dagger]=\delta_{nm}.
\end{equation}
Their redundancy allows the bosonic commutation relation to be rewritten as
\begin{equation}
\hat{\psi}_n^\dagger\hat{\psi}_n+\hat{\bar{\phi}}_n^\dagger\hat{\phi}_n=-1,
\end{equation}
indicating that the sum of quasiparticles and quasiholes is negative. As a result, although the ghost problem of first-quantized hole-like states is resolved by introducing the $\mathcal{CPT}$-inner product, the second-quantized quasihole Fock states necessarily involve negative quasiparticle excitations and thus appear as quasiparticle Fock states with negative norms. Thus, $\hat{H}$ should be presented separately in terms of quasiparticles and quasiholes:
\begin{equation}
\hat{H}=\sum_{n=1}^N\epsilon_n(\hat{\psi}_n^\dagger\hat{\psi}_n+\frac{1}{2})=\sum_{n=1}^N-\epsilon_n(\hat{\bar{\phi}}_n^\dagger\hat{\phi}_n+\frac{1}{2}).
\end{equation}
Although the opposite spectra of quasiparticles and quasiholes have been observed experimentally~\cite{E.Verhagen2022}, the ghost problems in the second quantization description of bosonic holes have not been recognized and solved in previous works. 


\subsection{$\mathcal{PT}$ symmetry broken regime}

The coupling between bosonic particles and holes induces spontaneous $\mathcal{PT}$-symmetry breaking, resulting in complex spectra ($\epsilon_n\notin \mathbb{R}$) that manifest as dynamical instability~\cite{B.Wu2001,B.Galilo2015,B.Wu2003,S.Burger2001,L.Fallani2004,O.Morsch2006,C.J.Wang2010,N.R.Bernier2014,I.A.Bhat2015,M.A.Khamehchi2017,H.Lyu2024}. In this regime, the $\mathcal{PT}$ symmetry of the eigenvectors is broken, so the conventional $\mathcal{PT}$ inner product is no longer well defined. To obtain the corrected inner product, we assume $\langle \psi_{nl}\vert$ as the left eigenvector of $\mathcal{H}$ with eigenvalue $\epsilon_n$. Then, we have  
\begin{equation}
{(\langle \psi_{nl}\vert \mathcal{H}\vert\psi_n\rangle)}^\dagger=\langle \psi_{n}\vert \mathcal{H}^\dagger\vert\psi_{nl}\rangle=\epsilon_n^\ast.
\end{equation}
It is clear that $\vert\psi_{nl}\rangle$ is the eigenvector of $\mathcal{H}^\dagger$ with eigenvalue $\epsilon_{n}^{\ast}$. Defining $\vert\psi_{n\ast}\rangle$ as the right eigenvector of $\mathcal{H}$ with eigenvalue $\epsilon_n^\ast$, Eq.~\eqref{Eq7} yields
\begin{equation}
\mathcal{H}^\dagger\tau_3\vert\psi_{n\ast}\rangle=\epsilon_n^\ast\tau_3\vert\psi_{n\ast}\rangle.
\end{equation}
Thus, $\langle\psi_{nl}\vert$ is given as
\begin{equation}
\langle\psi_{n\ast}\vert\tau_3={(\mathcal{PT}\vert\psi_{n\ast}\rangle)}^T,
\end{equation}
then the corrected $\mathcal{PT}$-inner product becomes 
\begin{equation}
{(\mathcal{PT}\vert\psi_{n\ast}\rangle)}^T\vert\psi_m\rangle=\langle \psi_{n\ast}\vert\tau_3\vert\psi_m\rangle=\delta_{nm}.
\end{equation} 

In this case, the interplay of PH symmetry and generalized $\mathcal{PT}$ symmetry still allow us to divide the eigenvalues into two groups $\{\epsilon_n,-\epsilon_n\}$, whose eigenvectors are denoted by $\{\vert\psi_n\rangle, \vert\tilde{\psi}_{n\ast}\rangle=C\vert\psi_{n\ast}\rangle\}$. Similarly, the corrected $\mathcal{PT}$-inner products of the latter $\vert\tilde{\psi}_{n\ast}\rangle$ are negative:
\begin{equation}
{({(\mathcal{PT}\vert\tilde\psi_{m}\rangle)}^T\vert\tilde\psi_{n\ast}\rangle)}^\dagger=\langle \psi_{n\ast}\vert-\mathcal{P}\vert\psi_{m}\rangle=-\delta_{nm}.
\end{equation}
Counterintuitively, although the $\mathcal{PT}$ symmetry of the eigenvectors is broken, one can still define a $\mathcal{C}$-symmetry operator of the form
\begin{equation}
\mathcal{C}=\vert\psi_{n}\rangle\langle\psi_{n\ast}\vert\mathcal{P}+\vert\tilde{\psi}_{n\ast}\rangle\langle\tilde{\psi}_{n}\vert\mathcal{P},
\end{equation}
labeling the sigh of $\mathcal{PT}$ norms 
\begin{equation}
    \mathcal{C}\vert\psi_n\rangle=\vert\psi_n\rangle, \quad \mathcal{C}\vert\tilde{\psi}_{n\ast}\rangle=-\vert\tilde{\psi}_{n\ast}\rangle.
\end{equation}
Such that, the corrected $\mathcal{CPT}$ inner products could always possess positive norms
\begin{equation}
   {(\mathcal{PT}\vert\psi_{n\ast}\rangle)}^T\mathcal{C}\vert\psi_n\rangle={(\mathcal{PT}\vert\tilde{\psi}_{n}\rangle)}^T\mathcal{C}\vert\tilde{\psi}_{n\ast}\rangle=1. 
\end{equation}
Inserting the identity
\begin{equation}\mathbb{I}_{2N}=\sum_{n=1}^{N}\vert\psi_n\rangle\langle\psi_{n\ast}\vert\tau_3-\vert\tilde{\psi}_{n\ast}\rangle\langle\tilde{\psi}_n\vert\tau_3,
\end{equation}
$\hat{H}$ can be diagonalized as 
\begin{equation}
\hat{H}=\sum_{n=1}^N\epsilon_n(\hat{\psi}_n^\dagger\hat{\psi}_{n\ast}+\frac{1}{2})=\sum_{n=1}^N-\epsilon_n(\hat{\bar{\phi}}_{n\ast}^\dagger\hat{\phi}_n+\frac{1}{2}),
\end{equation}
where $\hat{\psi}_{n\ast}=\langle\psi_{n\ast}\vert\tau_3\hat{\alpha}=-\hat{\bar{\phi}}_{n\ast}^\dagger$, fulfilling $[\hat{\psi}_{n\ast},\hat{\psi}_m^\dagger]=[\hat{\phi}_n,\hat{\bar{\phi}}_{m\ast}^\dagger]=\delta_{nm}$. 

Such complex spectra, and the resulting dynamical instability induced by the coupling between bosonic particles and holes, have been observed in inhomogeneous BECs~\cite{S.Burger2001,L.Fallani2004} and in optomechanical systems~\cite{N.R.Bernier2018,E.Verhagen2022}. The observation of quasihole branches, together with complex spectra, has confirmed the physical existence of bosonic holes. However, the Fock states of bosonic holes and their physical interpretation remain absent from existing literature, leaving the description and understanding of the system in the presence of holes still inaccessible.

\begin{figure}[tpb]
    \centering
    \includegraphics[width=0.75\linewidth]{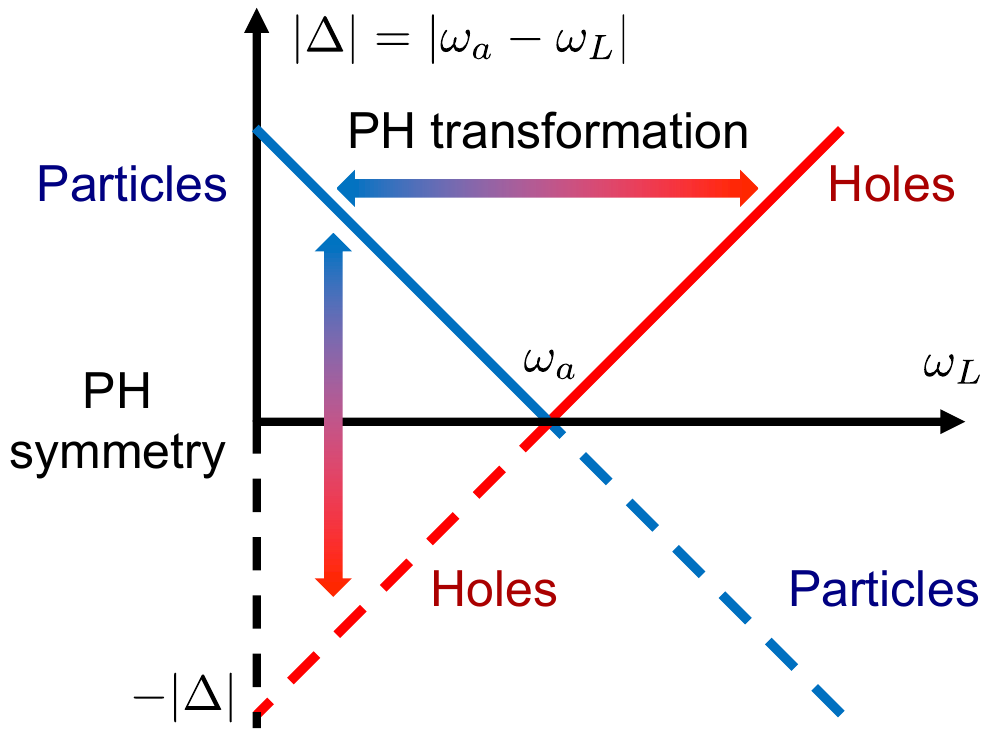}
    \caption{Bosonic excitation spectrum in QBSs relative to the frequency of a driving field $\omega_L$ (`Fermi level'), where $\pm\Delta$ are the relative frequencies of particles and holes, respectively, with $\omega_a$ being the intrinsic frequency of the particles.}
    \label{fig1}
\end{figure}

\section{Fock space and concept of bosonic holes}

\subsection{Particle-hole transformation.}

To address the ghost problem of bosonic holes, the key lies in identifying hole Fock states with positive norms. The challenge is that their vacuum needs to be annihilated by $\hat{a}^\dagger$, an idea that has been mentioned in several works~\cite{C.M.Bender2008,B.Holdom2024} but fails to be realized in a conventional Hilbert space, where bosonic occupations are unbounded. Here, we construct the hole Fock states via a PH transformation generated by a non-unitary but Hermitian operator, 
\begin{equation}
\hat{\Omega}=e^{i\frac{\pi}{4}(e^{-i\theta}\hat{a}^2-e^{i\theta}\hat{a}^{\dagger 2})}.
\end{equation}

A mathematically PH transformation for bosons first appeared in a similar form in Quantum Theory of Finite Systems by Blaizot and Ripka~\cite{J.-P.Blaizot1986}, where it was derived using a “quasi-spin” formalism in analogy with the fermionic PH transformation. However, it should be noted that because neither the energy spectrum of bosonic hole bands nor the complex spectrum or dynamical instability of QBSs had been observed at that time, the authors regarded such a transformation as lacking clear physical relevance. Consequently, it received little attention in the following decades. 

Here, we show that the PH transformation can be generated from the dynamics of a single-mode dissipative pairing interaction. The two-mode version of the dissipative paring interaction has been theoretically proposed in recent years and is considered to be physically realizable~\cite{A.A.Clerk2023}. We further show that the bosonic PH transformation also leads to clear physical applications: 
it establishes an exact mapping between the positive- and negative-particle-occupation space and enables to describe the Fock states of bosonic holes with positive norms in the particle representation. 

We begin with the non-Hermitian single-mode dissipative pairing (SDP) Hamiltonian
\begin{equation}
\hat{H}_{\rm SDP}=g(e^{i\theta}\hat{a}^{\dagger 2}-e^{-i\theta}\hat{a}^2)/2.
\end{equation}
A remarkable feature of this many-particle Hamiltonian is that its BdG Hamiltonian $\mathcal{H}_{\rm SDP}$ governing its elementary excitations is Hermitian. It can be obtained from the Heisenberg equation of motion 
\begin{equation}
i\frac{d}{dt}\left(\begin{array}{cc}
\hat{a} \\
\hat{a}^\dagger 
\end{array}\right)=\left(\begin{array}{cc}
0 & ge^{i\theta}\\
ge^{-i\theta}& 0 
\end{array}\right)\left(\begin{array}{cc}
\hat{a} \\
\hat{a}^\dagger
\end{array}\right)=\mathcal{H}_{\rm SDP}\left(\begin{array}{cc}
\hat{a} \\
\hat{a}^\dagger
\end{array}\right)\label{18}.
\end{equation}
Then, $\hat{H}_{\rm SDP}$ can be rewritten as 
\begin{equation}
    \hat{H}_{\rm SDP}=1/2\hat{\bar{\alpha}}^\dagger\mathcal{H}_{\rm SDP}\hat{\alpha}\label{19},
\end{equation}
where $\hat{\alpha}={(\hat{a},\hat{a}^\dagger)}^T$ and $\hat{\bar{\alpha}}^\dagger=(\hat{a}^\dagger,-\hat{a})$. 

Note that real spectra of non-Hermitian many-particle Hamiltonians are usually attributed to pseudo-Hermiticity (generalized $\mathcal{PT}$ symmetry) of their single-particle Hamiltonians. However,  Eq.~\ref{19} shows that the PH space can provide an alternative way to realize real spectra for a non-Hermitian many-particle Hamiltonian, and even restore Hermiticity in the BdG representation. The Hermitian BdG Hamiltonian brings unitary and periodic evolution in the PH space, and therefore, the dynamics represent coherent PH conversion. The field operators at time $t$ can be obtained from Eq.~\ref{18} as
\begin{equation}
\begin{split}
\hat{a}(t)=&\hat{a}(0)\cos(gt)+\hat{h}(0)\sin(gt),\\
\hat{a}^\dagger(t)=&\hat{a}^\dagger(0)\cos(gt)+\hat{\bar{h}}^\dagger(0)\sin(gt),
\end{split}
\end{equation}
where $\hat{h}(0)\equiv-ie^{i\theta}\hat{a}^\dagger(0)$ and $\hat{\bar{h}}^\dagger(0)\equiv-ie^{-i\theta}\hat{a}(0)$, fulfilling $[\hat{h}(0),\hat{\bar{h}}^\dagger(0)]=1$. 

Choosing $t=\pi/(2g)$ yields a complete particle-hole transformation:
\begin{equation}
\begin{split}
\hat{a}(\pi/2g)=&-ie^{i\theta}\hat{a}^\dagger(0)=\hat{h}(0),\\
\hat{a}^\dagger(\pi/2g)=&-ie^{-i\theta}\hat{a}(0)=\hat{\bar{h}}^\dagger(0).
\end{split}
\end{equation}
Thus, the PH transformation can be defined as 
\begin{equation}
\hat{\Omega}=e^{-i\hat{H}_{\rm SDP}\pi/(2g)}=e^{i\frac{\pi}{4}(e^{-i\theta}\hat{a}^2-e^{i\theta}\hat{a}^{\dagger 2})},
\end{equation}
which gives rise to 
\begin{equation}
\hat{\Omega}^{-1}\hat{a}\hat{\Omega}=-i\hat{a}^\dagger e^{i\theta}=\hat{h}, \quad \hat{\Omega}^{-1}\hat{a}^\dagger\hat{\Omega}=-i\hat{a}e^{-i\theta}=\hat{\bar{h}}^\dagger.
\end{equation}
The phase factor $e^{i\theta}$ comes from the phase of the pairing interaction in $\hat{H}_{\rm SDP}$ and can be gauged away by $\hat{a}^\dagger e^{i\theta/2}\to \hat{a}^\dagger$ and $\hat{a}e^{-i\theta/2}\to \hat{a}$. 

It is worth to emphasize that the PH symmetry of the BdG Hamiltonian mentioned earlier is conceptually distinct from the PH transformation of operators discussed here. To illustrate this difference, we consider the free bosonic Hamiltonian $\hat{H}=\omega(\hat{a}^\dagger\hat{a}+1/2)$. The former PH symmetry originates from PH redundancy, that is, particle modes can be rewritten into hole modes with opposite frequencies, 
\begin{equation}
    \hat{H}=\omega(\hat{a}^\dagger\hat{a}+1/2)=-\omega(\hat{\bar{h}}^\dagger\hat{h}+1/2).
\end{equation}
In contrast, the latter corresponds to an explicit PH transformation of operators, which maps particle modes onto hole modes with the same frequencies
\begin{equation}
\hat{\Omega}^{-1}\hat{H}\hat{\Omega}=\omega(\hat{\bar{h}}^\dagger\hat{h}+1/2),
\end{equation}
as illustrated in Fig.~\ref{fig1}.


\subsection{Fock space of bosonic holes.}

The transformation between particle and hole operators in the Heisenberg picture induces a corresponding transformation of particle and hole Fock states in the Schrodinger picture. Using the PH transformation, the hole vacuum is defined as 
\begin{equation}
\vert0\rangle_h=\hat{\Omega}^{-1}\vert0\rangle_a
\end{equation}
to ensure 
\begin{equation}
\hat{h}\vert0\rangle_h=-i\hat{a}^\dagger e^{i\theta}\vert0\rangle_h=\hat{\Omega}^{-1}\hat{a}\hat{\Omega}\hat{\Omega}^{-1}\vert0\rangle_{a}=0.
\end{equation}
Particularly, this state is biorthogonal due to the non-unitarity of $\hat{\Omega}$, so the hole states are not located in the conventional Hilbert space. 

Taking $\hat{h}$ ($\hat{\bar{h}}$) and $\hat{\bar{h}}^\dag$ ($\hat{h}^\dagger$) as the annihilation and creation operators of the right (left) eigenvectors, the biorthogonal Fock states of the hole are given by 
\begin{equation}
|n\rangle_{h}={(\hat{\bar{h}}^{\dagger})}^{n}/ \sqrt{n!}|0\rangle_{h}=\hat{\Omega}^{-1}|n\rangle_a,
\end{equation} 
and 
\begin{equation}
|m\rangle_{\bar{h}}={(\hat{h}^{\dagger})}^{m}/ \sqrt{m!}|0\rangle_{\bar{h}}=\hat{\Omega}|m\rangle_a,
\end{equation}
where $\hat{\bar{h}}\vert0\rangle_{\bar{h}}=0$. These yield the Fock states of bosonic holes with positive norms 
\begin{equation}{}_{\bar{h}}\langle m|n\rangle_{h}=\delta_{mn}
\end{equation}
and occupied by negative particle numbers 
\begin{equation}
_{\bar{h}}\langle n\vert\hat{a}^\dagger\hat{a}\vert n\rangle_h=_{\bar{h}}\langle n\vert-\hat{h}\hat{\bar{h}}^\dagger\vert n\rangle_h=-(n+1).
\end{equation}


\subsection{`Fermi surface' and `Fermi level' of bosonic holes}

Negative particle occupation can be interpreted by introducing a mean-field background, in which positive and negative excitations are defined relative to a macrooccupation mean particle number, functioning as a `Fermi surface.' We bring it with a displacement operator 
\begin{equation}
\hat{D}(\bar{a})=e^{\bar{a}\hat{a}^\dagger-\bar{a}^\ast\hat{a}}.
\end{equation}
In the displaced space, $n$ bosonic holes correspond to removing $n+1$ particles from the mean number $\bar{n}=\vert\bar{a}\vert^2$, i.e.,
\begin{equation}
    {}_{\bar{h}}\langle n\vert \hat{D}^{-1}(\bar{a})\hat{a}^\dagger\hat{a}\hat{D}(\bar{a})\vert n\rangle_{h}=\vert\bar{a}\vert^2-(n+1)\label{47}.
\end{equation}

Such macrooccupation naturally occurs in experimental systems that support a hole branch or exhibit complex spectra and dynamical instabilities induced by the coupling of bosonic particles and holes, such as inhomogeneous BECs~\cite{B.Wu2001} and coherently driven optomechanical systems~\cite{N.R.Bernier2018,E.Verhagen2022}. In the latter case, the coherent driving is described by
\begin{equation}
\hat{H}_{\rm D}=\Delta\hat{a}^\dagger\hat{a}+\lambda(\hat{a}^\dagger+\hat{a}),
\end{equation}
where $\Delta$ denotes the detuning between the cavity mode and driving field, and $\lambda$ is the driving strength. The system evolves according to the Lindblad master equation
\begin{equation}
    \frac{d\hat{\rho}}{dt}=-i[\hat{H}_{\rm D},\hat{\rho}]+\frac{\gamma}{2}\{2\hat{z}\hat{\rho}\hat{z}^\dagger-\hat{z}^\dagger\hat{z}\hat{\rho}-\hat{\rho}\hat{z}^\dagger\hat{z}\}\label{6},
\end{equation}
where $\hat{z}=\hat{a}$ ($\hat{a}^\dagger$) denotes the incoherent loss (pump). As shown in Appendix A, their steady-state solutions, i.e., $d\hat{\rho}_{ss}/dt=0$, are
\begin{equation}
\hat{\rho}_{ss}=\hat{D}(\bar{a})\vert0\rangle_{a}{}_{a}\langle0\vert\hat{D}^{-1}(\bar{a}), \quad \bar{a}=\frac{-i\lambda}{\gamma/2+i\Delta},
\end{equation}
and
\begin{equation}
\hat{\rho}_{ss}=\hat{D}(\bar{a})\vert0\rangle_{h}{}_{h}\langle0\vert\hat{D}^{-1}(\bar{a}), \quad \bar{a}=\frac{i\lambda}{\gamma/2-i\Delta},
\end{equation}
for incoherent loss and pump, respectively. Thus, the displaced particle vacuum emerges from the balance between coherent driving and incoherent loss~\cite{G.S.Agarwal1988}, and the displaced hole vacuum may arise from the competition between the coherent driving and incoherent pump processes. 

Notably, particle and hole frequencies become relative to the driving-field frequency, whereas in BECs they are given with respect to the effective chemical potential. These frequency reference frames can be regarded as the bosonic analogue of the `Fermi level.' The negative particle frequency, usually associated with dynamical instabilities~\cite{N.R.Bernier2014}, can occur when the driving-field frequency exceeds the particle frequency in quantum optical systems~\cite{N.R.Bernier2018} or when the chemical potential is modulated by periodic potentials in inhomogeneous BECs~\cite{B.Wu2001}. 

By focusing on physics near the `Fermi surface' and referenced to the `Fermi level,'  hole degrees of freedom facilitate rigorous studies of elementary excitations in bosonic systems without the problems of ghosts, negative particle excitations, and negative energies. More broadly, these problems have long existed in higher-derivative quantum field theory, extending beyond QBSs, where bosonic hole theory may also yield new understanding~\cite{S.W.Hawking2002,C.M.Bender2008,P.D.Mannheim2005}. 


\section{Particle-hole duality}

In Hermitian QBSs, the generalized $\mathcal{PT}$ symmetry originates from the inherent PH-symmetric structure of bosonic Bogoliubov-de-Gennes (BdG) Hamiltonians~\cite{J.-P.Blaizot1986,S.Lieu2018,L.Viola2020}. Because particles and holes are redundant, half of the eigenstates acquire positive norms (positive $\mathcal{C}$-parity), while the other half carry negative norms (negative $\mathcal{C}$-parity). This logic is reversed in non-Hermitian systems:  $\mathcal{PT}$ symmetry was discovered first~\cite{C.M.Bender1998}, and the hidden $\mathcal{C}$ symmetry was later revealed by the equal numbers of positive- and negative-$\mathcal{PT}$-norm states~\cite{C.M.Bender2002}, interpreted as ``particle'' and ``anti-particle'' states, respectively~\cite{C.M.Bender2015}. 

However, without knowing the basis, one cannot distinguish the BdG Hamiltonian of a Hermitian system from the single-particle Hamiltonian of a non-Hermitian system. This ambiguity has attracted great interest in recent years, for instance, using the bosonic BdG Hamiltonian to study $\mathcal{PT}$ dynamics without dissipation~\cite{A.A.Clerk2019}. Alternatively, it may also be impossible to distinguish the BdG Hamiltonian of a non-Hermitian system from the single-particle Hamiltonian of a Hermitian system, such as the dissipative pairing Hamiltonian mentioned earlier.

All of these demonstrate that although these systems have different Hermiticity, they describe the same physics at the single-particle level in different PH bases. Therefore, they can be connected by the local PH transformation,
\begin{equation}
\hat{\Omega}_{i}^{-1}\hat{H}\hat{\Omega}_i=\frac{1}{2}\hat{\Omega}_{i}^{-1}\hat{\bar{\alpha}}_i^\dagger\mathcal{H}\hat{\alpha}_i\hat{\Omega}_i=\frac{1}{2}\hat{\bar{\beta}}_i^\dagger\mathcal{H}\hat{\beta}_i=\hat{\tilde{H}},
\end{equation} 
where 
\begin{equation}
\hat{\beta}_i=\hat{\Omega}_i^{-1}\hat{\alpha}\hat{\Omega}_i=(\hat{a}_1\cdots-i\hat{a}_i^\dagger e^{i\theta}\cdots \hat{a}_N,\hat{a}_1^\dagger\cdots-i\hat{a}e^{-i\theta}\cdots \hat{a}_N^\dagger)^T
\end{equation}
and 
\begin{equation}
\hat{\bar{\beta}}_i^\dagger=\hat{\Omega}_i^{-1}\hat{\bar{\alpha}}^\dagger\hat{\Omega}_i
\end{equation}
with $[\hat{\beta}_i,\hat{\bar{\beta}}_i^\dagger]=I_{2N}$. For every eigenmode $\hat{\psi}_j$ of $\hat{H}$ with Fock states $\vert n\rangle_{\psi_j}$, there exists an eigenmode $\hat{\tilde{\psi}}_j$ in $\tilde{H}$, denoted by 
\begin{equation}
\hat{\Omega}_i^{-1}\hat{\psi}_j\hat{\Omega}_i.
\end{equation}
This eigenmode has the same frequency as $\hat{\psi}_j$ and its Fock states can be given by 
\begin{equation}
\vert n\rangle_{\tilde{\psi}_j}=\hat{\Omega}^{-1}_i\vert n\rangle_{\psi_j}.
\end{equation}
The PH transformation preserves the locality structure and spectral properties of the original Hamiltonian, and the transformed states and the original states represent essentially the same physics in different bases, implying the PH duality~\cite{E.Cobanera2010,D.X.Nguyen2017}. 

Moreover, the full dynamical information about the evolution of dual states under the dual Hamiltonian is encoded in the initial system. For example, if $\vert\psi(0)\rangle$ and $\vert\psi(t)\rangle$ are the initial state and final state of the original system, then the dynamics under the dual system reads
\begin{equation}
    e^{-i\hat{\tilde{H}}t}\hat{\Omega}_i^{-1}\vert\psi(0)\rangle=\hat{\Omega}_{i}^{-1}e^{-i\hat{H}t}\vert\psi(0)\rangle=\hat{\Omega}_{i}^{-1}\vert\psi(t)\rangle.\label{10}
\end{equation}
This PH duality is nontrivial because the solution can be well-known before the PH transformation, but still hidden after. In addition, such a non-unitary duality can connect the Hermitian and non-Hermitian QBSs and apply to any regime beyond the unitary duality~\cite{F.J.Dyson1956,M.Znojil2022,V.P.Flynn2020}.


\section{Hidden hole degrees of freedom in non-Hermitian QBSs}

We have demonstrated the PH duality between Hermitian and non-Hermitian QBSs. For non-Hermitian QBSs with $\mathcal{PT}$-symmetric single-particle Hamiltonians, states with opposite $\mathcal{C}$-parity may appear independent of bosonic holes, as they come from system-environment coupling rather than PH redundancy. Below, we adopt the standard $\mathcal{CPT}$ approach and use the PH duality to reveal the hidden hole degrees of freedom in such systems through a simple and experimentally feasible example: the dissipative beamsplitter Hamiltonian~\cite{C.M.Hu2018,wcyu2019,C.M.Hu2020,M.H.Yung2020},
\begin{equation}
\hat{H}_{\rm DBS}=\Delta_{1}\hat{a}_1^{\dagger}\hat{a}_1+\Delta_{2}\hat{a}_2^{\dagger}\hat{a}_2+ig(\hat{a}_1\hat{a}_2^\dagger + \hat{a}_1^{\dagger}\hat{a}_2)\label{58}.
\end{equation}
Defining $\hat{\beta}={(\hat{a}_1, \hat{a}_2)}^T$, $\hat{H}_{\rm DBS}$ can be rewritten as 
\begin{equation}
    \hat{H}_{\rm DBS}=\hat{\beta}^\dagger\mathcal{H}_{\rm SPH}\beta,
\end{equation}
where
\begin{equation}
\mathcal{H}_{\rm SPH}=\left(\begin{array}{cc}
\Delta_1& ig\\
ig& \Delta_2
\end{array}\right)
\end{equation}
called the single-particle Hamiltonian, describing the spectrum of single-particle excitations.

When $\Delta_1=-\Delta_2$, the single-particle Hamiltonian reads
\begin{equation}
\mathcal{H}_{\rm APT}=\left(\begin{array}{cc}
\Delta_1& ig\\
ig& -\Delta_1
\end{array}\right).
\end{equation}
In general cases, one commonly define $\mathcal{P}=\sigma_1$, under which $\{\mathcal{H}_{\rm APT}, \mathcal{PT}\}=0$, and $\mathcal{H}_{\rm APT}$ is therefore referred to as an $\mathcal{APT}$-symmetric Hamiltonian~\cite{F.Yang2017,Y.Yang2020}. However, $\mathcal{H}_{\rm APT}$ also exhibit a generalized $\mathcal{PT}$ symmetry, satisfying $[\mathcal{H}_{\rm APT}, \mathcal{PT}]=0$ with $\mathcal{P}=\sigma_3$.

For $\Delta_1>g$, $\mathcal{H}_{\rm APT}$ has real eigenvalues 
\begin{equation}
\omega_{\pm}=\pm\sqrt{\Delta_1^2-g^2},
\end{equation}
with $\mathcal{PT}$-unbroken ``particle'' and ``anti-particle'' states exhibiting opposite $\mathcal{PT}$-norms 
\begin{equation}
{(\mathcal{P}\mathcal{T}\vert\psi_\pm\rangle)}^T \vert\psi_\pm\rangle=\langle\psi_\pm\vert\tau_3\vert\psi_\pm\rangle=\pm1,
\end{equation}
and opposite $\mathcal{C}$-parity $\mathcal{C}\vert\psi_\pm\rangle=\pm\vert\psi_\pm\rangle$. The charge-conjugation operator $\mathcal{C}$ commute with $\mathcal{PT}$, then both states possess positive $\mathcal{CPT}$-norms, i.e., 
\begin{equation}
{(\mathcal{P}\mathcal{T}\vert\psi_\pm\rangle)}^T\mathcal{C}\vert\psi_\pm\rangle=\langle\psi_\pm\vert\pm\tau_3\vert\psi_\pm\rangle=1.
\end{equation}
Expanding $\mathcal{H}_{\rm APT}$ in its eigenvectors
\begin{equation}
\mathcal{H}_{\rm APT}=\omega_+\vert\psi_+\rangle\langle\psi_+\vert\sigma_3+\omega_-\vert\psi_-\rangle\langle\psi_-\vert(-\sigma_3),
\end{equation}
the eigenmodes of 
\begin{equation}
\hat{H}_{\rm DBS}=\hat{\beta}^\dagger \mathcal{H}_{\rm APT}\hat{\beta}
\end{equation}
read 
\begin{equation}
    \begin{split}
        \hat{\psi}_+=&\,\langle\psi_{+}\vert +\sigma_3 \hat{\beta}=u\hat{a}_{1}+ iv\hat{a}_{2}, \\
        \hat{\psi}_-=&\,\langle\psi_-\vert -\sigma_3 \hat{\beta}=-iv\hat{a}_1+u\hat{a}_2, \\
        \hat{\bar{\psi}}_{+}^\dagger=&\hat{\beta}^\dagger \vert \psi_{+}\rangle=u\hat{a}_1^\dagger+iv\hat{a}_2^\dagger,\\
        \hat{\bar{\psi}}_{-}^\dagger=&\hat{\beta}^\dagger \vert \psi_{-}\rangle=-iv\hat{a}_1^\dagger+u\hat{a}_2^\dagger\label{67}, 
    \end{split}
\end{equation} 
where $\hat{\beta}={(\hat{a}_1, \hat{a}_2)}^T$ and $u^2-v^2=1$, satisfying $[\hat{\psi}_\pm,\hat{\bar{\psi}}_\pm^\dagger]=1$.

Since $\hat{\beta}$ contains no hole degree of freedom and the eigenmodes appear to be superpositions of particles, i.e., $\hat{a}_1$ and $\hat{a}_2$ in $\hat{\psi}_{\pm}$ annihilate the particle vacuum $\vert0\rangle_{a_1}$ and $\vert0\rangle_{a_2}$, respectively. It is therefore natural to postulate that the ground state of the system is the tensor product of two particle vacuums. However, under this assumption, we find that the ghosts still appear as components in second-quantized states, even though the ghost problem in the single-particle level has been solved by $\mathcal{CPT}$ theory, e.g.,
\begin{equation}
    \begin{split}
\langle0\vert\hat{\psi}_+\hat{\bar{\psi}}_+^\dagger\vert0\rangle=&u^2{}_{a_1}\langle1\vert1\rangle_{a_1}-v^2{}_{a_2}\langle1\vert1\rangle_{a_2}=1,\\
\langle0\vert\hat{\psi}_-\hat{\bar{\psi}}_-^\dagger\vert0\rangle=&-v^2{}_{a_1}\langle1\vert1\rangle_{a_1}+u^2{}_{a_2}\langle1\vert1\rangle_{a_2}=1\label{68}.
    \end{split}
\end{equation}

\begin{figure}[tbp]
\includegraphics[width=0.95\linewidth]{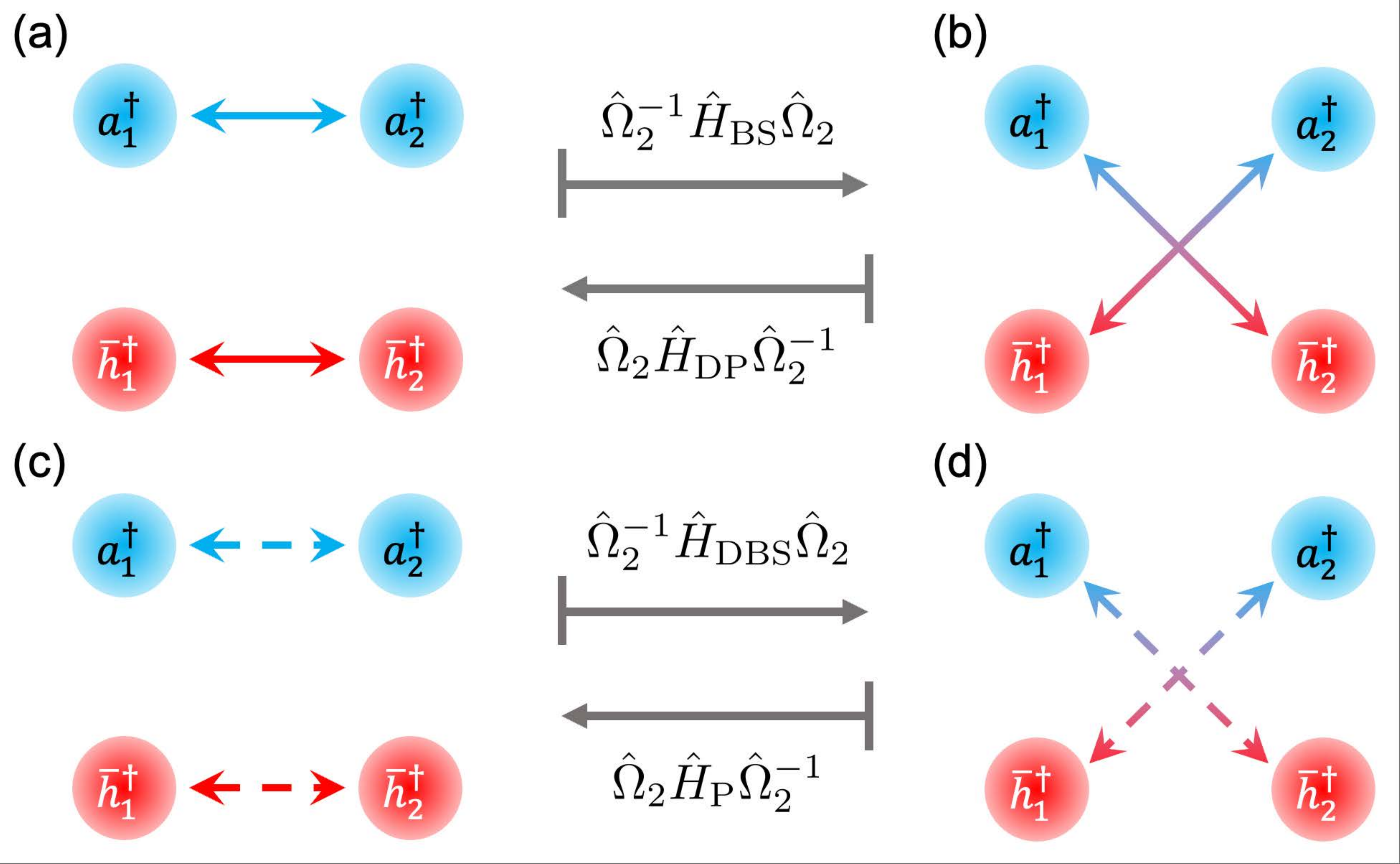}
\caption{Basic forms of two-mode QBSs, (a) $\hat{H}_{\rm BS}$, (b) $\hat{H}_{\rm DP}$, (c) $\hat{H}_{\rm DBS}$, and (d) $\hat{H}_{\rm P}$, where the solid and dashed lines denote coherent and dissipative couplings, respectively. (a)[(c)] shows $\hat{H}_{\rm BS}$ ($\hat{H}_{\rm DBS}$) with coherent (dissipative) particle-particle and hole-hole conversions, while (b)[(d)] shows $\hat{H}_{\rm DP}$ ($\hat{H}_{\rm P}$) with coherent (dissipative) PH conversions.}
\label{fig2}
\end{figure}

Such ghosts with negative particle expectations imply the hole degrees of freedom in the quasiparticles of non-Hermitian $\mathcal{PT}$ symmetric systems. The opposite signs of particle-component expectations for the two eigenmodes originate from the opposite $\mathcal{C}$-parity, reflecting that these eigenmodes belong to PH conjugate spaces.

To clarify this point, we invoke the PH duality introduced in the previous section. We find that $\hat{H}_{\rm DBS}$ is intimately related to another well-known model
\begin{equation}
    \hat{H}_{\rm P}=\Delta_{1}\hat{a}_1^{\dagger}\hat{a}_1-\Delta_{2}\hat{a}_2^{\dagger}\hat{a}_2+g(\hat{a}_1^\dagger \hat{a}_2^\dagger + \hat{a}_1\hat{a}_2)\label{69},
\end{equation}
which describes the coherent creation and annihilation of particles in pairs. Although $\hat{H}_{\rm P}$ describes an entirely different system from $\hat{H}_{\rm DBS}$—the latter representing environmentally induced dissipative particle-particle conversions—the two Hamiltonians are related by a local PH transformation 
\begin{equation}
    \hat{\Omega}_2^{-1}\hat{H}_{\rm DBS}\hat{\Omega}_2=\hat{H}_{\rm P},
\end{equation}
where $\hat{\Omega}_2=e^{i\frac{\pi}{4}(\hat{a}_2^2-\hat{a}_2^{\dagger 2})}$. This PH duality implies that the two models encode the same physics in different PH representations. 

At the same parameter $\Delta_1=-\Delta_2$, $\hat{H}_{\rm P}$ reduces to the mean-field description of weakly interacting Bose gases. Its quasiparticles are the famous Bogoliubov modes
\begin{equation}
\hat{B}_{ij}=u\hat{a}_{i}+v\hat{a}_{j}^\dagger~(i,j=1,2\quad i\ne j)
\end{equation}
with a degenerate spectrum $\omega_+=\sqrt{\Delta_1^2-g^2}$, forming the foundation of superfluidity~\cite{N.Bogoliubov1971}. 

Based on the PH duality, the annihilation operators of quasiparticles for $\hat{H}_{\rm DBS}$ are given by 
\begin{equation}
\begin{split}
\hat{\Omega}_2\hat{B}_{12}\hat{\Omega}_2^{-1}=&u\hat{a}_1-v\hat{\bar{h}}_2^\dagger=u\hat{a}_{1}+iv\hat{a}_2=\hat{\psi}_+,\\
\hat{\Omega}_2\hat{B}_{21}\hat{\Omega}_2^{-1}=&v\hat{a}_1^\dagger-u\hat{h}_2=v\hat{a}_1^\dagger+iu\hat{a}_2^\dagger=i\hat{\bar{\psi}}_-^\dagger,
\end{split}
\end{equation}
with the degenerate spectrum $\omega_+$ as $\hat{H}_{\rm P}$. This result confirms our earlier analysis of Eq.~\ref{68}, i.e., even in non-Hermitian systems, single-particle eigenstates with opposite $\mathcal{C}$-parity yield eigenmodes residing in PH conjugate spaces. Therefore, to obtain quasiparticles in the same space and thereby describe the ground state of the system (the quasiparticle vacuum), we should regard either $e^{i\phi}\hat{\bar{\psi}}_-^\dagger$ or $e^{i\phi}\hat{\bar{\psi}}_+^\dagger$ in Eq.~\ref{67} as an annihilation operator, where $e^{i\phi}$ is a phase that can be gauge away. The resulting single-particle spectra of $\hat{H}_{\rm DBS}$ are no longer the generally believed $\omega_\pm$, but become degenerate $\omega_+$ or $\omega_-$.

In the field of quantum optics, the Bogoliubov modes are usually described by the squeezed operator~\cite{C.C.Gerry2005,Y.D.Wang2013}
\begin{equation}
\hat{S}_{a_1a_2}(r)=\exp[r\hat{a}_1\hat{a}_2-H.c.]
\end{equation}
as
\begin{equation}
  \hat{B}_{ij}=\hat{S}_{a_1a_2}(r)\hat{a}_{i}\hat{S}_{a_1a_2}^{-1}(r)  
\end{equation}
Then the ground state of the Bogoliubov modes is the two-mode squeezed vacuum
\begin{equation}
\vert G\rangle_{\rm P}=\hat{S}_{a_1a_2}(r)\vert00\rangle_{a_1a_2}
\end{equation}
with entanglement $E_N=2\vert r\vert$. These suggest that we can introduce the PH two-mode squeezed operator
\begin{equation}
\hat{\Omega}_2\hat{S}_{a_1a_2}\hat{\Omega}_2^{-1}=\hat{S}_{a_1\bar{h}_2}(r)=\exp[r\hat{a}_1\hat{\bar{h}}_2-r^\ast\hat{a}_1^\dagger\hat{h}_2^\dagger],
\end{equation}
such that the quasiparticles of $\hat{H}_{\rm DBS}$ can be described as
\begin{equation}
\hat{\psi}_+/i\hat{\bar{\psi}}_-^\dagger=\hat{\Omega}_2\hat{B}_{12}/\hat{B}_{21}\hat{\Omega}_2^{-1}=\hat{S}_{a_1\bar{h}_2}(r)\hat{a}_1/\hat{\bar{h}}_2\hat{S}_{a_1\bar{h}_2}^{-1}(r),
\end{equation}
and the ground state of $\hat{H}_{\rm DBS}$ is the PH two-mode squeezed vacuum
\begin{equation}
    \vert G\rangle_{\rm DBS}=\hat{S}_{a_1\bar{h}_2}(r)\vert00\rangle_{a_1\bar{h}_2}=\hat{\Omega}_2\vert G\rangle_{\rm P}\label{78}.
\end{equation}

The introduction of bosonic holes and PH duality provides a fundamentally innovative perspective for understanding the real spectra of non‑Hermitian systems. First, when Eq.~\ref{78} is identified as the system ground state for discussing quasiparticle excitations, the negative-norm issue appearing in Eq.~\ref{68} no longer arises. Although excitations with negative particle numbers remain, this problem is naturally resolved when the quasiparticles are excitated near the `Fermi surface', as already noted in Sec.~\Rmnum{4}. 

A significant motivation for studying $\mathcal{CPT}$ quantum mechanics stems from the $\mathcal{CPT}$ theorem of relativistic local quantum field theory. It is therefore of considerable interest to seek a nonrelativistic analogue of this theorem. Unlike the charge-conjugation operator in quantum field theory, however, the $\mathcal{C}$ operator in $\mathcal{CPT}$ quantum mechanics depends on the choice of Hamiltonian and is thus generally not regarded as the direct analogue of relativistic charge conjugation. For QBSs, by contrast, the structure of the BdG Hamiltonian is dominated by PH redundancy, and therefore determines the forms of the $\mathcal{P}$ symmetry and the $\mathcal{C}$ operator. Since holes in nonrelativistic condensed-matter systems play a role analogous to antiparticles in relativistic quantum field theory, the $\mathcal{C}$ operator in QBSs, which distinguishes quasiparticles residing in PH conjugate spaces, may be viewed as the condensed-matter analogue of the charge-conjugation operator.

Second, PH duality reveals that the same Hamiltonian admits distinct physical interpretations in different PH representations. In particular, the PH duality and the PH two‑mode squeezed vacuum ground state indicate that, within the real spectrum regime, the interaction described by $\hat{H}_{\rm DBS}$ is more appropriately viewed as the creation and annihilation of PH excitations in pairs
\begin{equation}
    \hat{H}_{\rm DBS}=\Delta_1\hat{a}_1^\dagger\hat{a}_1-\Delta_2\hat{h}_2^\dagger\hat{\bar{h}}_2+g(\hat{a}_1\hat{\bar{h}}_2+\hat{a}_1^\dagger\hat{h}_2^\dagger)~\label{79}
\end{equation}
rather than as dissipative particle‑particle conversion in Eq. \ref{58}.

Another significant and experimentally feasible example illustrating the hole degrees of freedom in non‑Hermitian QBSs is provided by the dissipative pairing Hamiltonian discussed earlier in the context of the PH transformation. Here we consider the two-mode case~\cite{A.A.Clerk2023}
\begin{equation}
    \hat{H}_{\rm DP}=\Delta_{1}\hat{a}_1^{\dagger}\hat{a}_1-\Delta_{2}\hat{a}_2^{\dagger}\hat{a}_2-ig(\hat{a}_1^\dagger \hat{a}_2^\dagger + \hat{a}_1\hat{a}_2),
\end{equation} 
Unlike the preceding examples, the real spectrum of the non‑Hermitian $\hat{H}_{\rm DP}$ does not rely on unbroken PT symmetry. Instead, it originates from the Hermiticity of the BdG Hamiltonian endowed by the hole degree of freedom. 

Defining $\hat{\beta}={(\hat{a}_1, \hat{a}_2^\dagger)}^T$, the BdG Hamiltonian that describes the single-particle spectrum is governed by
\begin{equation}
    \hat{H}_{\rm DP}=\hat{\bar\beta}^\dagger\mathcal{H}_{\rm DP}\hat\beta,
\end{equation}
where
\begin{equation}
\mathcal{H}_{\rm DP}=\left(\begin{array}{cc}
\Delta_1& g\\
g& \Delta_2
\end{array}\right)
\end{equation}
and $\hat{\bar{\beta}}^\dagger=(\hat{a}_1^\dagger, -\hat{a}_2)$. The Hermiticity of $\mathcal{H}_{\rm DP}$ in the PH space indicates that it describes coherent PH conversion. 

This interpretation becomes transparent through the PH duality of $\hat{H}_{\rm DP}$ and the well-known beamsplitter Hamiltonian
\begin{equation}
    \hat{H}_{\rm BS}=\Delta_{1}\hat{a}_1^{\dagger}\hat{a}_1+\Delta_{2}\hat{a}_2^{\dagger}\hat{a}_2+g(\hat{a}_1\hat{a}_2^\dagger + \hat{a}_1^{\dagger}\hat{a}_2).
\end{equation}
As shown in Fig.~\ref{fig2} (a), $\hat{H}_{\rm BS}$ describes the coherent conversion between particles or holes. Within the PH duality, $\hat{H}_{\rm DP}$ exhibits the same physics as $\hat{H}_{\rm BS}$ in the PH representation 
\begin{equation}
    \hat{H}_{\rm DP}=\hat{\Omega}_2^{-1}\hat{H}_{\rm BS}\hat{\Omega}_2=\Delta_{1}\hat{a}_1^{\dagger}\hat{a}_1+\Delta_{2}\hat{h}_2\hat{\bar{h}}_2^\dagger+g(\hat{a}_1^\dagger \hat{h}_2 + \hat{a}_1\hat{\bar{h}}_2^\dagger),
\end{equation}
showing the coherent conversions between particles and holes, as shown in Fig.~\ref{fig2} (b). This conversion indicates that the system’s excitations reside near the direct-product state formed by the displaced particle vacuum of mode 1 and the displaced hole vacuum of mode 2. Coincidentally, in experimentally realizable setups, the two modes are inevitably subject to incoherent gain and dissipation, respectively~\cite{A.A.Clerk2023}. In light of the displaced particle vacuum and displaced hole vacuum discussed above, it is therefore reasonable to infer that the approximate steady state of $\hat{H}_{\rm DP}$ is precisely the direct-product state, as shown in Appendix B.

This mechanism, which gives rise to real spectra in non-Hermitian systems independently of $\mathcal{PT}$ symmetry, is highly unconventional. This insight points to a deeper underlying reason governing the emergence of real spectra in non‑Hermitian systems.


\section{Dynamics in particle-hole space and particle-hole entanglement}

In the existing literature, where well-defined hole states are absent, the dynamics of QBSs have been studied primarily either through the Heisenberg equations of motion or through state evolutions in the Hilbert space spanned by particle Fock states. Here, we show that neither approach provides a complete description of QBS dynamics, because the dynamics depend crucially on the space of initial states. First, state evolution in the Hilbert space spanned solely by particle Fock states cannot capture the PH dynamics. In the Heisenberg picture, the evolution of the field operators under $\hat{H}_{\rm DP}$ is governed by a unitary matrix
\begin{equation}
    i\frac{d}{dt}\hat{\beta}=\mathcal{H}_{\rm DP}\hat{\beta}\label{85},
\end{equation}
where $\hat{\beta}={(\hat{a}_1, \hat{a}_2^\dagger)}^T$, yielding a periodic coherent PH conversion. In contrast, in the Schrödinger picture, the non-unitary evolution operator $e^{-i\hat{H}_{\rm DP}t}$ acting on the orthogonal Hilbert space does not produce periodic evolution. However, when the dynamics are formulated in the biorthogonal PH space spanned by particle and hole Fock states, for example, 
\begin{equation}
\vert\psi(t)\rangle=C_1(t)\vert10\rangle_{a_1h_2}+C_2(t)\vert01\rangle_{a_1h_2},
\end{equation}
the evolution remains confined to the excitation-conservation subspace. The Schrödinger equation, 
\begin{equation}
    i\frac{d}{dt}\vert\psi(t)\rangle=\hat{H}_{\rm DP}\vert\psi(t)\rangle
\end{equation}
then gives
\begin{equation}
    i\frac{d}{dt}\left(\begin{array}{cc}
C_1(t)\\
C_2(t)
\end{array}\right)=\mathcal{H}_{\rm DP}\left(\begin{array}{cc}
C_1(t)\\
C_2(t)
\end{array}\right),
\end{equation}
thereby exhibiting unitary dynamics identical to those in the Heisenberg picture.

Second, space-dependent dynamics cannot be captured by the Heisenberg equations of motion because of the redundancy between particle and hole operators. Here we take the dual Hamiltonian $\hat{H}_{\rm DBS}$ and $\hat{H}_{\rm P}$ to illustrate this. We have demonstrated that $\hat{H}_{\rm DBS}$ should be interpreted as the creation and annihilation of PH pairs in the real spectrum regime. In the other perspective, the PH duality also suggests that the Hermitian $\hat{H}_{\rm P}$ can also be interpreted as dissipative PH conversion
\begin{equation}
    \hat{H}_{\rm P}=\Delta_1\hat{a}_1^\dagger\hat{a}_1+\Delta_2\hat{\bar{h}}_2^\dagger\hat{h}_2+ig(\hat{a}_1^\dagger\hat{h}_2+\hat{a}_1\hat{\bar{h}}_2^\dagger)\label{88},
\end{equation}
as demonstrated in Fig.~\ref{fig2} (d). This occurs when $\mathcal{PT}$ symmetry is spontaneously broken, and the quasiparticles of the system are no longer Bogoliubov modes. To see this clearly, we take $\Delta_1=\Delta_2$, where the quasiparticles of $\hat{H}_{\rm DBS}$ and $\hat{H}_{\rm P}$ are described by
\begin{equation}
\hat{\psi}_{\pm}=(\hat{a}_1\pm\hat{a}_2)/\sqrt{2},
\end{equation}
and
\begin{equation}
\hat{\phi}_{\pm}=\hat{\Omega}_2^{-1}\hat{\psi}_{\pm}\hat{\Omega}_2=(\hat{a}_1\pm\hat{h}_2)/\sqrt{2},
\end{equation}
respectively, with the same frequencies $\omega_\pm=\Delta_1\pm ig$. It is clear that the eigenspace of $\hat{H}_{\rm DBS}$ is spanned by the particle Fock states, while the eigenspace of $\hat{H}_{\rm P}$ is spanned by the particle and hole Fock states.

In the Heisenberg picture, the redundancy between particle and hole operators implies that the two representations of $\hat{H}_{\rm DBS}$ (as shown in Eq.~\ref{58} and Eq.~\ref{79}) and $\hat{H}_{\rm P}$ (as shown in Eq.~\ref{69} and Eq.~\ref{88}) obey the same equations of motion, namely,
\begin{equation}
    i\frac{d}{dt}\left(\begin{array}{cc}
\hat{a}_1\\
\hat{a}_2
\end{array}\right)=i\frac{d}{dt}\left(\begin{array}{cc}
\hat{a}_1\\
i\hat{\bar{h}}_2^\dagger
\end{array}\right)=\left(\begin{array}{cc}
\Delta_1 & ig\\
ig & \Delta_1
\end{array}\right)\left(\begin{array}{cc}
\hat{a}_1\\
\hat{a}_2
\end{array}\right),
\end{equation}
and
\begin{equation}
    i\frac{d}{dt}\left(\begin{array}{cc}
\hat{a}_1\\
\hat{h}_2
\end{array}\right)=i\frac{d}{dt}\left(\begin{array}{cc}
\hat{a}_1\\
-i\hat{a}_2^\dagger
\end{array}\right)=\left(\begin{array}{cc}
\Delta_1 & ig\\
ig & \Delta_1
\end{array}\right)\left(\begin{array}{cc}
\hat{a}_1\\
\hat{h}_2
\end{array}\right),
\end{equation}
respectively.

However, in the Schrödinger picture, because the particle and hole Fock spaces are not equivalent, the two representations of $\hat{H}_{\rm DBS}$ and $\hat{H}_{\rm P}$ enable one to study the dynamics of the system in two distinct state spaces. As we show below, the resulting dynamics depend sensitively on whether the initial state space coincides with or differs from the eigenspace.

We first consider the case in which the initial state and the eigenstates belong to distinct spaces. A typical example is the evolution of the particle vacuum, or of coherent states, under $\hat{H}_{\rm P}$~\cite{C.C.Gerry2005,S.W.Du2022}. In this case, the initial state cannot be decomposed into a superposition of eigenstates, leading to unbounded evolution
\begin{equation}
    e^{-i\hat{H}_{\rm P}t}\vert00\rangle_{a_1a_2}=\hat{S}_{a_1a_2}(r)\vert00\rangle_{a_1a_2},
\end{equation}
where $\hat{S}_{a_1a_2}(r)\vert00\rangle_{a_1a_2}$ is a two-mode squeezed vacuum with a squeezing parameter $r=-igt$ growing linearly in time. Although the eigenstates of $\hat{H}_{\rm P}$ are PH superpositions, the effect of $\hat{H}_{\rm P}$ depends on the space of the initial state on which it acts; here it manifests itself as pair creation or annihilation of particles.

It is readily seen that, because the eigenstates of Hermitian QBSs in the complex energy spectrum are PH superpositions, the time evolution of certain observables becomes unbounded and non‑periodic when a generic initial state in the particle Fock space is considered. Thus, this regime is commonly referred to as dynamically unstable.

Moreover, the PH duality suggests that analogous dynamically unstable behavior can also occur in particle-conserving non-Hermitian QBSs when the initial state is taken in PH space, e.g., 
\begin{equation}
\hat{\Omega}_2\vert00\rangle_{a_1a_2}=\vert00\rangle_{a_1\bar{h}_2}.
\end{equation}
The evolution of this state under $\hat{H}_{\rm DBS}$ can be obtain via $\hat{H}_{\rm DBS}=\hat{\Omega}_2\hat{H}_{\rm P}\hat{\Omega}_2^{-1}$, namely, 
\begin{equation}
    e^{-i\hat{H}_{\rm DBS}t}\vert00\rangle_{a_1\bar{h}_2}=\hat{\Omega}_2\hat{S}_{a_1a_2}(r)\vert00\rangle_{a_1a_2}=\hat{S}_{a_1\bar{h}_2}(r)\vert00\rangle_{a_1\bar{h}_2}.
\end{equation}
Similarly, although the eigenstates of $\hat{H}_{\rm DBS}$ are superpositions of particle states, when the initial state is in PH space, $\hat{H}_{\rm DBS}$ manifests itself as pair creation and annihilation of PH excitations, leading to unbounded evolution and the generation of infinite PH entanglement.

We next consider the case that the initial states and eigenstates belong to the same space, for example, a single-particle excited state $\vert10\rangle_{a_1a_2}$ evolving under $\hat{H}_{\rm DBS}$~\cite{M.H.Yung2020}. Because they belong to the same space, the initial state admits a decomposition into the eigenstates 
\begin{equation}
\vert10\rangle_{a_1a_2}=(\hat{\psi}_++\hat{\psi}_-)/\sqrt{2}\vert00\rangle_{a_1a_2}.
\end{equation}
The evolution is therefore confined to the single-particle subspace, and the steady state is the eigenstate with the largest imaginary part 
\begin{equation}
    \hat{\psi}_+^\dagger\vert00\rangle_{a_1a_2}=(\vert10\rangle_{a_1a_2}+\vert01\rangle_{a_1a_2})/\sqrt{2}\vert00\rangle_{a_1a_2},
\end{equation}
which is a Bell state. The PH duality implies that the steady PH Bell state with finite PH entanglement
\begin{equation}
\hat{\Omega}_2^{-1}\hat{\psi}_+^\dagger\vert00\rangle_{a_1a_2}=\hat{\bar{\phi}}_+^\dagger\vert00\rangle_{a_1h_2}=(\vert10\rangle_{a_1h_2}+\vert01\rangle_{a_1h_2})/\sqrt{2} 
\end{equation}
can be generated by the evolution of the PH single-excited state 
\begin{equation}
\Omega_2^{-1}\vert00\rangle_{a_1a_2}=\vert10\rangle_{a_1h_2}=(\hat{\bar{\phi}}_-^\dagger+\hat{\bar{\phi}}_+^\dagger)/\sqrt{2}\vert00\rangle_{a_1h_2}
\end{equation}
under $\hat{H}_{\rm P}$. Therefore, PH states provide a feasible approach to studying steady-state solutions of Hermitian QBSs in the dynamically unstable regime. In this case, the time evolution of certain observables remains bounded and eventually reaches a stable value.

These examples show that the hole Fock space introduced here enables access to a broader class of dynamical phenomena than is possible within previous approaches. More importantly, the relation between the initial-state space and the system eigenspace provides a clear explanation for the emergence of bounded and unbounded evolution in both Hermitian and non-Hermitian QBSs in the complex spectrum regime.


\section{Chiral flows of particle-hole excitations}

With the $\mathcal{PT}$ symmetric BdG Hamiltonian, the non-Hermitian AB effect has been proposed and experimentally observed in Hermitian QBSs~\cite{E.Verhagen2022}. In turn, the Hermiticity of non-Hermitian QBSs reflected by PH duality allows us to investigate the Hermitian AB effect in PH spaces.

\begin{figure}[tpb]
    \centering
    \includegraphics[width=1.\linewidth]{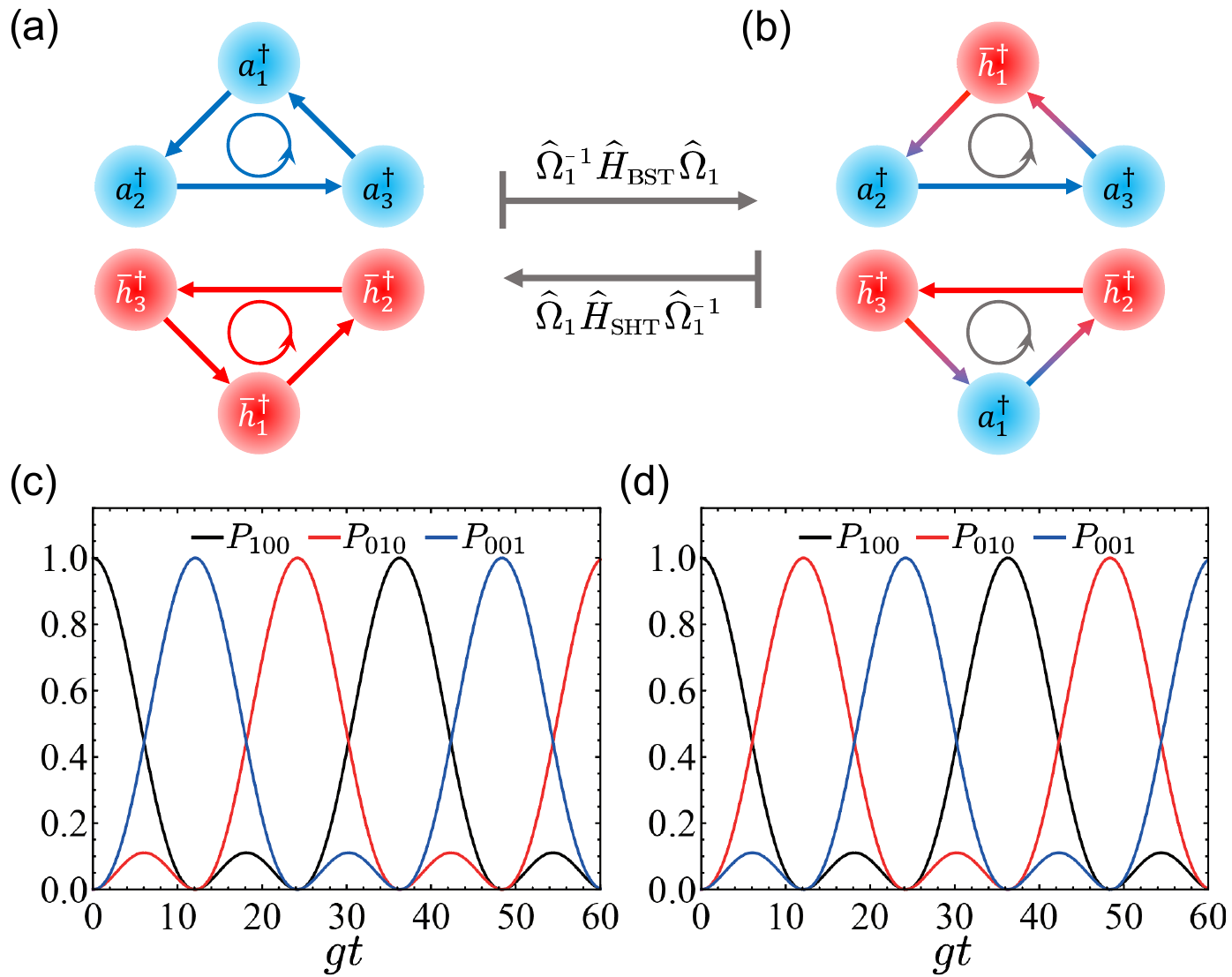}
    \caption{Schematics of chiral flows in the trimers with the Hamiltonians (a) $\hat{H}_{BST}$ and (b) $\hat{H}_{SHT}$, when $\Phi=-\frac{\pi}{2}$. (c) and (d) plot the population evolutions of single-excited states in each loop when $\Phi=\pi/2$ and $\Phi=-\pi/2$, respectively, where $P_{100}=\vert C_1(t)\vert^2$, $P_{010}=\vert C_2(t)\vert^2$ and $P_{001}=\vert C_3(t)\vert^2$.}
    \label{fig3}
\end{figure}

We consider a ring network composed of three bosonic modes, as depicted in Fig.~\ref{fig3} (a) and (b). In the particle-conserving case, this system is referred to as a `beamsplitter trimer', described by the Hamiltonian 
\begin{equation}
\hat{H}_{\rm BST}=\sum_{ i=1,j\ne i}^{3}ge^{i\varphi_{ji}}\hat{a}_i^{\dagger}\hat{a}_j,
\end{equation}
with phases satisfying $\varphi_{ji}=-\varphi_{ij}$~\cite{E.Verhagen2022,J.Martinis2017}. Applying a local PH transformation on mode 1 yields its non-Hermitian dual version, the `single-hole trimer,' described by
\begin{equation}
    \hat{H}_{\rm SHT}=\sum_{ i=2,j\ne i}^{3}ge^{i\varphi_{ji}}\hat{a}_i^{\dagger}\hat{a}_j+g(e^{i\varphi_{1i}}\hat{a}_i^\dagger\hat{h}_1+e^{i\varphi_{i1}}\hat{\bar{h}}_1^\dagger\hat{a}_i).
\end{equation}
The restoration of excitation-conservation of $\hat{H}_{\rm SHT}$ reflects its invariance under the U(1) gauge transformation in the PH space, 
\begin{equation}
\hat{U}_{\rm PH}=e^{i\varphi(\hat{\bar{h}}_1^\dagger\hat{h}_1+\sum_{i=2}^3\hat{a}_i^\dagger\hat{a}_i)}.
\end{equation}
The effective AB phase 
\begin{equation}
\Phi=\varphi_{12}+\varphi_{23}+\varphi_{31}
\end{equation}
acts the synthetic flux threading the loops and is invariant under local gauge transformations, i.e., $\hat{a}_i^\dagger\to \hat{a}_i^\dagger e^{i\varphi}$ and $\hat{\bar{h}}_i^\dagger\to \hat{\bar{h}}_i^\dagger e^{i\varphi}$ for particle and hole nodes, respectively. In the symmetric gauge $\varphi_{12}=\varphi_{23}=\varphi_{31}=\Phi/3$, translational invariance permits diagonalization in discrete momentum bases
\begin{equation}
\begin{split}
&\hat{a}_{k_{BST}}=\sum_{j=1}^3e^{-i2\pi kj/3}\hat{a}_{j}/\sqrt{3},\\
&\hat{a}_{k_{SHT}}=[e^{-i2\pi k/3}\hat{h}_{1}+\sum_{j=2}^3e^{-i2\pi kj/3}\hat{a}_{j}]/\sqrt{3},
\end{split}
\end{equation}
for $k=\{-1,0,1\}$, with eigenfrequencies $\epsilon_k=2g\cos [(\Phi-2\pi k]/3)$. 

At trivial values fluxes $\Phi=z\pi~(z\in \mathbb{Z})$, the spectrum features twofold degeneracy, preserving time-reversal symmetry~\cite{J.Koch2010}. Conversely, non-trivial fluxes will break time-reversal symmetry via AB interference, which is the underlying mechanism for the chirality of quantum Hall edge states~\cite{T.Ozawa2019} and nonreciprocal transport~\cite{A.Metelmann2015}.

We now show chiral energy transport in the single excitation subspace, e.g., 
\begin{equation}
e^{-i\hat{H}_{\rm SHT}t}\vert\psi(0)\rangle_{h_1a_2a_3}=C_1(t)\hat{\bar{h}}_1^\dagger+\sum_{i=2}^3C_i(t)\hat{a}_i^\dagger\vert000\rangle_{h_1a_2a_3},
\end{equation}
with the initial coefficients $C_1(0)=1$, $C_2(0)=C_3(0)=0$. While the time evolution under $\hat{H}_{\rm SHT}$ is governed by the non-unitary operator $e^{-i\hat{H}_{\rm SHT}t}$, it generates unitary dynamics on biorthogonal PH states. The state evolution exhibits counterclockwise circulation, i.e., 
\begin{equation}
\vert100\rangle_{h_1a_2a_3}\rightarrow\vert010\rangle_{h_1a_2a_3}\rightarrow\vert001\rangle_{h_1a_2a_3}\rightarrow\vert100\rangle_{h_1a_2a_3}
\end{equation}
in the single-hole loop. When the flux is tuned to $\Phi=\pi/2$, the circulation direction of single-excited states is reversed, as shown in Fig.~\ref{fig3} (c) and (d). These chiral excitation circulations within PH spaces reveal an unexplored pathway for studying bosonic analogs of the quantum Hall effect in non-Hermitian QBSs.


\section{Conclusions}

In summary, we 
develop a second-quantization theory of bosonic holes and find the correspondence between PH conjugate spaces of eigenmodes and the $\mathcal{C}$-parity of single-particle eigenvectors in QBSs. Additionally, we unveil the PH duality, showing that the stable PH superposition, forbidden in Hermitian QBSs, can emerge in non-Hermitian QBSs. Our work not only paves the way for comprehensive investigations of QBSs, but also provides insights into charge-conjugation symmetry in $\mathcal{CPT}$ quantum mechanics as well as the real energy spectra in non-Hermitian systems.

Our results open up promising directions for future studies: characterizing the exact ground state of QBSs toward non-Hermitian topological classification~\cite{S.Lieu2018,Z.X.Zhou2020,X.-Y.Lv2023}, exploring the topologically protected Majorana bosons-superpositions of bosonic particles and holes-distinct from the bosonic Majorana zero modes~\cite{C.-E.Bardyn2012,A.McDonald2018,Q.-R.Xu2020,V.P.Flynn2021,E.Verhagen2024,A.Y.Kitaev2001}, and extending bosonic hole theory to higher-derivative quantum field theories and other fields with ghost problems in second-quantized states~\cite{B.Holdom2024,C.M.Bender2008}.

\section*{Acknowledgments}

We thank Q. Niu, P. Rabl and Z. Yan for stimulating discussions. This work was supported by the National Natural Science Foundation of China (Grant No. 12375025 and No. 11874432).

\section*{APPENDIX A: Generation of displaced hole vacuum}
Here, we detail the generation of the displaced particle vacuum (coherent state) and the displaced hole vacuum in the balance between coherent driving and incoherent loss or pump. We demonstrate it by considering a single-mode Hamiltonian with coherent driving, described as
\begin{equation}
    \hat{H}_{\rm D}=\Delta \hat{a}^\dagger\hat{a}+\lambda(\hat{a}^\dagger+\hat{a}),
\end{equation}
where $\Delta=\omega_a-\omega_d$ denotes the detuning of the system mode and driving field, and $\lambda$ represents the strength of the driving field. The dynamics of the system with incoherent loss can be described by a Lindblad master equation
\begin{equation}
 \frac{d\hat{\rho}}{dt}=-i[\Delta \hat{a}^\dagger\hat{a}+\lambda(\hat{a}^\dagger+\hat{a}),\hat{\rho}]+\frac{\gamma}{2}(2\hat{a}\hat{\rho}\hat{a}^\dagger-\hat{a}^\dagger\hat{a}\hat{\rho}-\hat{\rho}\hat{a}^\dagger\hat{a}),\label{S29}
\end{equation}
where $\hat{\rho}$ denotes the system density matrix. We introduce the new operator $\hat{b}=\hat{a}+\frac{i\lambda}{\gamma/2+i\Delta}$, so that the master equation Eq. (\ref{S29}) can be rewritten as 
\begin{equation}
 \frac{d\hat{\rho}}{dt}=-i[\Delta \hat{b}^\dagger\hat{b},\hat{\rho}]+\frac{\gamma}{2}(2\hat{b}\hat{\rho}\hat{b}^\dagger-\hat{b}^\dagger\hat{b}\hat{\rho}-\hat{\rho}\hat{b}^\dagger\hat{b}).
\end{equation}
In the steady state we have $\frac{d}{dt}\hat{\rho}_{ss}=0$ and thus it follows that we must have $\hat{b}\hat{\rho}_{ss}=\hat{\rho}_{ss}\hat{b}^\dagger=0$, i.e.,
\begin{equation}
(\hat{a}+\frac{i\lambda}{\gamma/2+i\Delta})\hat{\rho}_{ss}=\hat{\rho}_{ss}(\hat{a}^\dagger-\frac{i\lambda}{\gamma/2-i\Delta})=0,\label{S31}
\end{equation}
which means that the steady state is an eigenstate of the annihilation operator. Eq. (\ref{S31}) can be solved by
\begin{equation}
\hat{\rho}_{ss}=\hat{D}(\bar{a})\vert0\rangle_{a}{}_{a}\langle0\vert\hat{D}^{-1}(\bar{a})=\vert\bar{a}\rangle_a{}_{a}\langle\bar{a}\vert, \quad \bar{a}=\frac{-i\lambda}{\gamma/2+i\Delta},
\end{equation}
where $\vert \bar{a}\rangle$ is a coherent state. Thus, the competition between the coherent driving and the incoherent loss leads to a displaced particle vacuum in the steady-state regime~\cite{G.S.Agarwal1988}. We now consider the dynamics of the system with an incoherent pump, described as
\begin{equation}
 \frac{d\hat{\rho}}{dt}=-i[\Delta \hat{a}^\dagger\hat{a}+\lambda(\hat{a}^\dagger+\hat{a}),\hat{\rho}]+\frac{\gamma}{2}(2\hat{a}^\dagger\hat{\rho}\hat{a}-\hat{a}\hat{a}^\dagger\hat{\rho}-\hat{\rho}\hat{a}\hat{a}^\dagger).\label{S33}
\end{equation}
Similarly, we introduce the new operator $\hat{b}^\dagger=\hat{a}^\dagger+\frac{i\lambda}{\gamma/2+i\Delta}$, so that Eq. (\ref{S33}) can be rewritten as
\begin{equation}
 \frac{d\hat{\rho}}{dt}=-i[\Delta \hat{b}\hat{b}^\dagger,\hat{\rho}]+\frac{\gamma}{2}(2\hat{b}^\dagger\hat{\rho}\hat{b}-\hat{b}\hat{b}^\dagger\hat{\rho}-\hat{\rho}\hat{b}\hat{b}^\dagger).
\end{equation}
In the steady state, we have $\hat{b}^\dagger\hat{\rho}_{ss}=\hat{\rho}_{ss}\hat{b}=0$, i.e.,
\begin{equation}
(\hat{a}^\dagger+\frac{i\lambda}{\gamma/2+i\Delta})\hat{\rho}_{ss}=\hat{\rho}_{ss}(\hat{a}-\frac{i\lambda}{\gamma/2-i\Delta})=0,\label{S35}
\end{equation}
which means that the steady state is an eigenstate of the creation operator. Eq. (\ref{S35}) can be solved by 
\begin{equation}
\hat{\rho}_{ss}=\hat{D}(\bar{a})\vert0\rangle_{h}{}_{h}\langle0\vert\hat{D}^{-1}(\bar{a}), \quad \bar{a}=\frac{i\lambda}{\gamma/2-i\Delta}.
\end{equation}
Thus, the displaced hole vacuum may be generated by the competition between the coherent driving and the incoherent pump.

\section*{APPENDIX B: Steady particle-hole vacuum of two-mode dissipative pairing Hamiltonian}

Here we show that the approximate steady state of the physically realizable two-mode dissipative pairing Hamiltonian~\cite{A.A.Clerk2023}, which is inherently accompanied by incoherent loss and pump, is a displaced particle-hole vacuum under the strong coherent driving. We start with an effective two-mode dissipative pairing Hamiltonian
\begin{equation}
\hat{H}_{\rm DP}=\Delta_1\hat{a}_1^\dagger\hat{a}_1-\Delta_2\hat{a}_2^\dagger\hat{a}_2-ig(\hat{a}_1^\dagger\hat{a}_2^\dagger+\hat{a}_1\hat{a}_2),
\end{equation}
where $\Delta_1$ and $-\Delta_2$ are the frequencies relative to the local driving. Such an effective $\hat{H}_{\rm DP}$ can be obtained by considering the Lindblad master equation with jump operator $\hat{z}=u\hat{a}_1+v\hat{a}_2^\dagger$, described as
\begin{equation}
\begin{split}
    \frac{d\hat{\rho}}{dt}=&-i[\Delta_1\hat{a}_1^\dagger\hat{a}_1-\Delta_2\hat{a}_2^\dagger\hat{a}_2+\lambda_1(\hat{a}_1^\dagger+\hat{a}_1)+\lambda_2(\hat{a}_2^\dagger+\hat{a}_2),\hat{\rho}]\\
    &+\frac{\gamma}{2}\{2\hat{z}\hat{\rho}\hat{z}^\dagger-\hat{z}^\dagger\hat{z}\hat{\rho}-\hat{\rho}\hat{z}^\dagger\hat{z}\}~\label{128},
\end{split}
\end{equation}
where $\lambda_1$ and $\lambda_2$ are the strength of the local driving fields for $\hat{a}_1$ and $\hat{a}_2$. It is clear that the modes $\hat{a}_1$ and $\hat{a}_2$ are accompanied by local incoherent loss and pump, respectively. We introduce the operators $\hat{b}_1=\hat{a}_1+\frac{i\lambda_1}{u^2\gamma/2+i\Delta_1}$ and $\hat{b}_2^\dagger=\hat{a}_2^\dagger+\frac{i\lambda_2}{v^2\gamma/2-i\Delta_2}$, so that Eq. (\ref{128}) can be rewritten as
\begin{equation}
\begin{split}
    \frac{d\hat{\rho}}{dt}=&-i[\Delta_1\hat{b}_1^\dagger\hat{b}_1,\hat{\rho}]+\frac{u^2\gamma}{2}\{2\hat{b}_1\hat{\rho}\hat{b}_1^\dagger-\hat{b}_1^\dagger\hat{b}_1\hat{\rho}-\hat{\rho}\hat{b}_1^\dagger\hat{b}_1\}\\&-i[-\Delta_2\hat{b}_2\hat{b}_2^\dagger,\hat{\rho}]+\frac{v^2\gamma}{2}\{2\hat{b}_2^\dagger\hat{\rho}\hat{b}_2-\hat{b}_2\hat{b}_2^\dagger\hat{\rho}-\hat{\rho}\hat{b}_2\hat{b}_2^\dagger\}\\
    &+\frac{uv\gamma}{2}\{2\hat{a}_2^\dagger\hat{\rho}\hat{a}_1^\dagger+2\hat{a}_1\hat{\rho}\hat{a}_2-(\hat{a}_2\hat{a}_1+\hat{a}_1^\dagger\hat{a}_2^\dagger)\hat{\rho}\\
    &-\hat{\rho}(\hat{a}_2\hat{a}_1+\hat{a}_1^\dagger\hat{a}_2^\dagger)\}\label{129}.
\end{split}
\end{equation}
Neglecting the interaction terms in the third line in Eq. (\ref{129}), the approximate solution for $\frac{d}{dt}\hat{\rho}=0$ is given by 
\begin{equation}
\hat{\rho}_{ss}\approx\hat{D}_1(\bar{a}_1)\hat{D}_2(\bar{a}_2)\vert00\rangle_{a_1h_2}{}_{h_2a_1}\langle00\vert\hat{D}_2^{-1}(\bar{a}_2)\hat{D}_1^{-1}(\bar{a}_1)\label{130},
\end{equation}
where $\bar{a}_1=\frac{-i\lambda_1}{u^2\gamma/2+i\Delta_1}$ and $\bar{a}_2=\frac{i\lambda_2}{v^2\gamma/2+i\Delta_2}$. It means that the approximate steady state is a product state of the displaced particle vacuum of mode 1 and the displaced hole vacuum of mode 2. Substituting the approximate steady state solution Eq. (\ref{130}) into Eq. (\ref{129}), we obtain
\begin{equation}
\begin{split}
        \frac{d\hat{\rho}_{ss}}{dt}=&-\frac{uv\gamma}{2}\{\hat{D}_1(\bar{a}_1)\hat{D}_2(\bar{a}_2)(\bar{a}_1\hat{a}_2+\bar{a}_2^\ast\hat{a}_1^\dagger)\vert00\rangle_{a_1h_2}\\
        &{}_{h_2a_1}\langle00\vert\hat{D}_2^{-1}(\bar{a}_2)\hat{D}_1^{-1}(\bar{a}_1)\\
&+\hat{D}_1(\bar{a}_1)\hat{D}_2(\bar{a}_2)\vert00\rangle_{a_1h_2}\\&{}_{h_2a_1}\langle00\vert(\bar{a}_2\hat{a}_1+\bar{a}_1^\ast\hat{a}_2^\dagger)\hat{D}_2^{-1}(\bar{a}_2)\hat{D}_1^{-1}(\bar{a}_1)\},
\end{split}
\end{equation}
which indicates that near the approximate steady-state solution, only quantum jumps of mode-1 particles or mode-2 holes exist, with $\hat{a}_2=i\hat{\bar{h}}_2^\dagger e^{i\theta}$ and $\hat{a}_2^\dagger=i\hat{h}_2e^{-i\theta}$. Therefore, the system's states near the steady state may be a product state of the mode-1 displaced particle Fock state and the mode-2 displaced hole Fock state. For this reason, the effective two-mode dissipative pairing interaction $\hat{H}_{int}=-ig(\hat{a}_1^\dagger\hat{a}_2^\dagger+\hat{a}_1\hat{a}_2)$ obtained from \ref{129}, i.e., $-\frac{uv\gamma}{2}\{(\hat{a}_2\hat{a}_1+\hat{a}_1^\dagger\hat{a}_2^\dagger)\hat{\rho}-\hat{\rho}(\hat{a}_2\hat{a}_1+\hat{a}_1^\dagger\hat{a}_2^\dagger)\}=-i\{\hat{H}_{int}\hat{\rho}-\hat{\rho}\hat{H}_{int}^\dagger\}$ with $\frac{uv\gamma}{2}=g$, can be regarded as the coherent conversion between mode-1 particles and mode-2 holes for a given quantum trajectory. Note that the effect of the local driving and the incoherent loss/pump for mode 1 and mode 2 can be incorporated into the mean field. When considering the physics near the mean field, only the free Hamiltonian $\hat{H}_0=\Delta_1\hat{a}_1^\dagger\hat{a}_1-\Delta_2\hat{a}_2^\dagger\hat{a}_2$ and the interaction Hamiltonian $\hat{H}_{int}$ need to be taken into account, resulting in $\hat{H}_{\rm DP}=\hat{H}_0+\hat{H}_{int}$. By taking $v\to ie^{i\varphi}v$ in $\hat{z}$, the phase of the interaction can be modulated such that
\begin{equation}
\hat{H}_{\rm DP}=\Delta_1\hat{a}_1^\dagger\hat{a}_1-\Delta_2\hat{a}_2^\dagger\hat{a}_2+g(e^{i\varphi}\hat{a}_1^\dagger\hat{a}_2^\dagger-e^{-i\varphi}\hat{a}_1\hat{a}_2),
\end{equation}
which can be rewritten in terms of the particle operators of mode 1 and the hole operators of mode 2
\begin{equation}
\hat{H}_{\rm DP}=\Delta_1\hat{a}_1^\dagger\hat{a}_1+\Delta_2\hat{\bar{h}}_2^\dagger\hat{h}_2+ig(e^{i(\varphi-\theta)}\hat{a}_1^\dagger\hat{h}_2-e^{-i(\varphi-\theta)}\hat{a}_1\hat{\bar{h}}_2^\dagger),
\end{equation}
with $\hat{h}_2=-i\hat{a}_2^\dagger e^{i\theta}$ and $\hat{\bar{h}}_2^\dagger=-i\hat{a}_2e^{-i\theta}$.

\bibliographystyle{apsrev4-2}

\end{document}